\documentstyle[12pt,epsf]{article}

\newcommand{\beq}{\begin{equation}}
\newcommand{\eeq}{\end{equation}}
\newcommand{\bea}{\begin{eqnarray}}
\newcommand{\eea}{\end{eqnarray}}
\def\nn{\nonumber}

\newcommand{\tr}{\mbox{tr}}
\newcommand{\eqn}[1]{(\ref{#1})}
\newcommand{\quater}{{\bb H}} 
\newcommand{\quaters}{{\bbs H}} 
\newcommand{\complex}{{\bb C}} 
\newcommand{\complexs}{{\bbs C}} 
\newcommand{\zed}{{\bb Z}} 
\newcommand{\real}{{\bb R}} 
\newcommand{\reals}{{\bbs R}} 
\newcommand{\zeds}{{\bbs Z}} 
\newcommand{\ids}{{\bbs I}} 
\newcommand{\id}{{\bb I}} 
\newcommand{\klein}{{(-1)^{F_{\rm L}}}} 

\font\mybb=msbm10 at 12pt
\def\bb#1{\hbox{\mybb#1}}
\font\mybbs=msbm10 at 9pt
\def\bbs#1{\hbox{\mybbs#1}}

\newcommand{\newsection}[1]
{\vspace{4mm}
\pagebreak[3]
\addtocounter{section}{1}
\setcounter{equation}{0}
\setcounter{subsection}{0}
\begin{flushleft}
{\large\bf \thesection. #1}
\end{flushleft}
\nopagebreak
\medskip
\nopagebreak}

\newcommand{\newsubsection}[1]{
 \vspace{4mm}
\pagebreak[3]
\addtocounter{subsection}{1}
\noindent{ \bf \thesubsection. #1}
\nopagebreak
\vspace{1.7mm}
\nopagebreak}

\font\mybb=msbm10 at 12pt
\font\mybbs=msbm10 at 9pt

\makeatletter
\newdimen\normalarrayskip              
\newdimen\minarrayskip                 
\normalarrayskip\baselineskip
\minarrayskip\jot
\newif\ifold             \oldtrue            \def\new{\oldfalse}

\makeatother

\newlength{\extraspace}
\newlength{\extraspaces}
\begin{document}

\addtolength{\baselineskip}{.8mm}

\thispagestyle{empty}

\begin{flushright}
\baselineskip=12pt
NBI-HE-99-10\\
hep-th/9904153\\
\hfill{  }\\Revised Version, July 1999
\end{flushright}
\vspace{.5cm}

\begin{center}

\baselineskip=24pt

{\Large\bf{Brane Descent Relations in K-theory}}\\[15mm]

\baselineskip=12pt

{\bf Kasper Olsen} and {\bf Richard J. Szabo}
\\[5mm]

{\it The Niels Bohr Institute\\ Blegdamsvej 17, DK-2100\\ Copenhagen \O,
Denmark\\ {\tt kolsen , szabo @nbi.dk}}
\\[15mm]

\vskip 1 in

{\sc Abstract}

\begin{center}
\begin{minipage}{15cm}

The various descent and duality relations among BPS and non-BPS D-branes are
classified using topological K-theory. It is shown how the descent procedures
for producing type-II D-branes from brane-antibrane bound states by tachyon
condensation and $\klein$ projections arise as natural homomorphisms of
K-groups generating the brane charges. The transformations are generalized to
type-I theories and type-II orientifolds, from which the complete set of vacuum
manifolds and field contents for tachyon condensation is deduced. A new set of
internal descent relations is found which describes branes over orientifold
planes as topological defects in the worldvolumes of
brane-antibrane pairs on top of planes of higher dimension. The
periodicity properties of these relations are shown to be a consequence of the
fact that all fundamental bound state constructions and hence the complete
spectrum of brane charges are associated with the topological
solitons which classify the four Hopf fibrations.

\end{minipage}
\end{center}

\end{center}

\baselineskip=18pt

\noindent
\vfill
\newpage
\pagestyle{plain}
\setcounter{page}{1}

\newsection{Introduction}

One of the most interesting recent realizations about string solitons is that
their charges take values in terms of a generalized cohomology theory of their
Chan-Paton gauge bundles known as K-theory \cite{minmoore}--\cite{bgh}. The
original observation \cite{polchinski} that D-branes couple to spacetime
Ramond-Ramond (RR) fields suggests that their worldvolume topological charges
should be measured by ordinary cohomology classes, and it was long thought that
supersymmetry was necessary to ensure the stability of these solitonic
configurations. However, in other recent developments \cite{senbps}--\cite{bg}
it has been shown that the spectra of superstring theories can also contain
states which correspond to non-BPS solitons but which are nevertheless stable
because they carry conserved quantum numbers which prevents them from decaying
into the supersymmetric vacuum state. In this framework, D-branes are realized
as topological solitons coming from condensation of the tachyon field on the
worldvolumes of higher dimensional unstable configurations of branes. These
facts tie in nicely with the properties of the K-groups of a spacetime which
can be much more general than the corresponding cohomology groups. For
instance, the K-group can have torsion while the cohomology group is torsion
free \cite{minmoore}, lending a natural explanation to the fact that some
D-branes carry torsion charges \cite{senbps}.

The main property of K-theory which parallels the soliton constructions of
\cite{senbps}--\cite{bg} is its intimate relationship with homotopy theory. A
given category of fiber bundles always possesses a classifying space which
provides a universal bundle for it. The K-groups of the category are then
related to the homotopy groups of the classifying space. In some
instances it is possible to realize this equivalence in terms of stable
homotopy groups of finite dimensional homogeneous spaces. In this paper we will
exploit the stable homotopy properties of K-theory to describe the various
relationships that exist between stable and unstable string solitons. Such
connections between different types of D-branes are known as `descent
relations' \cite{witten,horava,sendescent} and they form a remarkable web of
mappings between BPS and non-BPS branes that provides various different ways of
thinking about the origins of D-branes. The situation for type-II
D-branes is depicted in fig. 1 \cite{sendescent}. If we consider,
say, a D$p$-brane anti-D$p$-brane (or D$\overline{p}$-brane) bound state pair
of type-IIB string theory ($p$ odd), then the spectrum contains a tachyonic
excitation whose ground state corresponds to the supersymmetric vacuum
configuration. However, one can consider instead a tachyonic kink solution on
the brane-antibrane pair which describes a non-BPS D$(p-1)$-brane of the IIB
theory. This system also contains a tachyonic excitation in its worldvolume
field theory, so that one can consider a tachyonic kink solution on the
D$(p-1)$-brane which results in a BPS D$(p-2)$-brane of IIB.

\begin{figure}[htb]
\epsfxsize=5in
\bigskip
\centerline{\epsffile{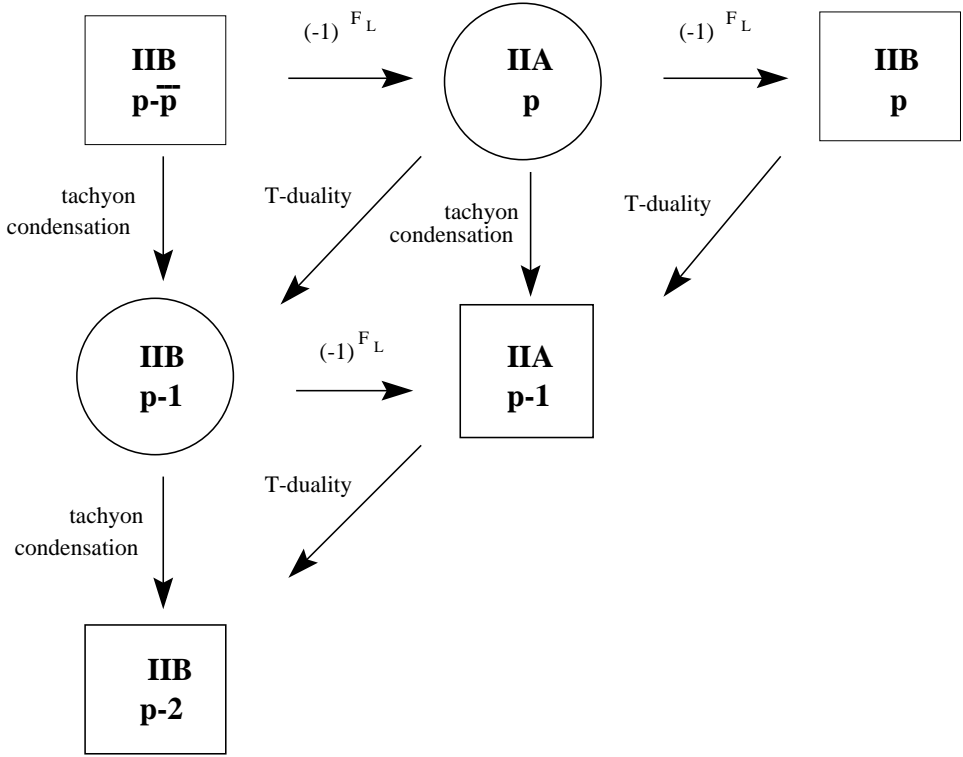}}
\caption{\baselineskip=12pt {\it The relationships between different D-branes
in type-II superstring theory. The squares represent stable supersymmetric BPS
branes or a combination of such a brane with its antibrane, while the circles
depict unstable non-BPS configurations. The horizontal arrows represent the
result of quotienting the theory by the operator $\klein$, the vertical arrows
the effect of constructing a tachyonic kink solution in the brane worldvolume
field theory, and the diagonal arrows the usual $T$-duality transformations.}}
\bigskip
\label{descentfig}\end{figure}

Another set of relations comes from modding out the $p$-$\overline{p}$ brane
pair by the operator $\klein$ which acts as $-1$ on all the Ramond sector
states in the left-moving part of the fundamental string worldsheet, and leaves
all other sectors unchanged. A careful study of the open string spectrum of
this configuration reveals that the result is a non-supersymmetric D$p$-brane
of IIA, and that a further quotient by $\klein$ yields a supersymmetric
$p$-brane of IIB \cite{sendescent,daspark}. When combined with the usual
$T$-duality transformations between the type IIB and IIA theories, we find that
any $p$-brane configuration in type-II superstring theory may be obtained from
a chain of higher dimensional brane configurations. In particular, all branes
of the type-II theories descend from a bound state of spacetime filling
D9-$\overline{\rm D9}$ pairs, which agrees with standard mathematical
constructions in K-theory \cite{witten,horava}.

The new understanding of the tachyon in an unstable brane configuration as a
Higgs type excitation in the spectrum of string states leads to a topological
classification of the resulting brane charges \cite{semz} when D-branes are
viewed as the tachyonic solitons. Generally, the topological charges of these
objects are determined by the homotopy groups of a homogeneous space $G/H$, where
$G$ is a compact Lie group and $H$ is a closed subgroup of $G$. The fibration
$H~{\buildrel i\over\hookrightarrow}~G~{\buildrel\pi\over\to}~G/H$, with $i$
the inclusion and $\pi$ the canonical projection, induces a long exact sequence
of homotopy groups,
\beq
\dots~\longrightarrow~\pi_{n-1}(H)~{\buildrel{i^*}\over\longrightarrow}~
\pi_{n-1}(G)~{\buildrel{\pi^*}\over\longrightarrow}~\pi_{n-1}(G/H)~{\buildrel
{\partial^*}\over\longrightarrow}~\pi_{n-2}(H)~\longrightarrow~\dots
\label{longexacthtpy}\eeq
In this paper we shall be primarily interested in two instances of this
homotopy sequence. The first one will be related to the weak Bott periodicity
theorem for the classical Lie groups \cite{bott}--\cite{bfn}, in which
case the induced boundary homomorphism $\partial^*$ is an isomorphism so that
$\pi_{n-1}(G/H)=\pi_{n-2}(H)$. In this case, $G$ is the worldvolume gauge group
of a given configuration of branes and the tachyon scalar field $T$ is a Higgs
field for the breaking of the gauge symmetry down to the subgroup $H$. The
tachyonic soliton must be accompanied by a worldvolume gauge field $A$ of
corresponding topological charge in the vacuum manifold $G/H$, in order that
the energy per unit worldvolume of the induced lower dimensional brane be
finite. A careful study of the tachyon potential \cite{horava,senbps} shows
that the brane worldvolume field theory contains finite energy, static soliton
solutions which have asymptotic pure gauge configurations at infinity,
\beq
T\simeq\lambda_0\,U~~~~~~,~~~~~~A\simeq U^{-1}\,dU
\label{puregauge}\eeq
where $\lambda_0$ is a constant, and $U$ is a $G/H$ valued function
corresponding to the identity map (of a given winding number) from the
asymptotic boundary of the worldvolume soliton to the group manifold of $G/H$.
If the induced brane configuration has codimension $n$ in the higher
dimensional worldvolume, then the corresponding soliton carries topological
charge taking values in $\pi_{n-1}(G/H)$.

Another instance of \eqn{longexacthtpy} that will be considered in the
following will involve the construction of K-theory classes based on the
properties of spinor modules. This mapping is known as the Atiyah-Bott-Shapiro
(ABS) construction \cite{karoubi,abs,spingeom} and it is equivalent to the
physical realization of a D-brane as the tachyonic soliton in the worldvolume
of a bound state system  of higher dimensional branes. It is based on an
intimate relationship between K-theory and the structure of Clifford algebras,
and it yields an explicit form for the classical tachyon field as Clifford
multiplication by a vector in the sphere bundle
\beq
SO(n)~\hookrightarrow~SO(n+1)~\longrightarrow~S^n
\label{spherebundle}\eeq
over the transverse space to the D-brane. This construction will be related to
the case where $\partial^*$ is a trivial mapping, i.e.
$\ker\partial^*=\pi_{n-1}(G/H)$, so that
$\pi_{n-1}(G/H)=\pi_{n-1}(G)/\pi_{n-1}(H)$. As we will see, one of the natural
consequences of this construction is that the induced D-brane is represented as
the topological soliton associated with the Hopf fibration \cite{husemoller}
\beq
S^{n-1}~\hookrightarrow~S^{2n-1}~\longrightarrow~S^n
\label{hopffib}\eeq
There are only four non-trivial Hopf fibrations, for $n=1,2,4,8$, which
correspond respectively to the real, complex, quaternion and octonion division
algebras over the field of real numbers. The classifying map of the fibration
is a generator of
\beq
\pi_{n-1}\Bigl(spin(n)\Bigr)\,/\,\pi_{n-1}\Bigl(spin(n-1)\Bigr)~
=~\left\{\new{\begin{array}{cll}\zed_2~~~~&,&~~n=1\\\zed~~~~&,
&~~n=2,4,8\end{array}}\right.
\label{hopfclassmap}\eeq
As we will show, the Hopf fibrations determine all the fundamental bound
state constructions of branes in type-I and type-II superstring theories, and
hence the complete spectrum of D-brane charges rests on the
fact that there are only four such bundles. For $n\neq 1$, the topological
charge of the corresponding soliton is given by the Pontryagin number density
which is proportional to $\tr(F^{n/2})$, where $F$ is the curvature of the
associated topologically non-trivial gauge field configuration. For $n=1$ the
charge is determined by a $\zed_2$-valued Wilson line, as in the standard
construction of non-BPS branes in type-I string theory \cite{senbps}. This fact
therefore also realizes all D-branes in terms of more conventional solitons,
such as the Dirac monopole and the $SU(2)$ Yang-Mills instanton, and it
moreover determines the explicit forms of the non-trivial gauge fields living
on the brane worldvolumes.

In this paper we will give a systematic and exhaustive mathematical
classification of the D-brane descent relations based on the periodicity
theorems of topological K-theory. In addition to providing many deeper insights
into the existing properties of branes within their K-theory/bound state
constructions, this analysis will also predict various new descent relations
and symmetries among D-branes. All facts about
K-theory that are used are explained throughout the paper. We will consider
only the simplest case of a flat ten dimensional spacetime manifold and
topologically trivial worldvolume embeddings, leaving the analysis of curved
manifolds as an interesting and important future generalization of the present
description (global aspects of the bound state constructions are discussed in
\cite{witten,garcia}).

In section 2 we will analyse the type-II theories, mainly to introduce the
relevant notations, constructions and ideas that will be used later on in the
case of more involved brane configurations. We begin by analysing the weak Bott
periodicity theorem for the stable homotopy groups of unitary symmetric spaces
using the ABS construction and the standard proof of Bott periodicity based on
the structure of Clifford algebra representations \cite{bfn}. We describe in
detail the descent relations among various branes in this formalism, and relate
them to the analysis of \cite{seninst}. We then proceed to show that these
natural isomorphisms are nothing but the $T$-duality transformations between
the IIB and IIA superstring theories. The K-theoretic realization of
$T$-duality
has been discussed in \cite{hori}--\cite{bgh}. For completeness, we shall give
some further insight into the structure of the $T$-duality maps in complex
K-theory, in a slightly different spirit than the previous analyses by
emphasizing the transformation properties of the worldvolume fields. We also
show how the
$\klein$ transformations are realized as natural homomorphisms in terms of
equivariant K-theory\footnote{Our analysis of the $\klein$ projections differs
from the results of \cite{gukov}.} and show that the mapping between the IIA
and IIB theories under the $\klein$ projection is formally the same as their
relationship under $T$-duality. We compare this K-theory construction with a
boundary state formalism \cite{daspark} which demonstrates that the $\klein$
quotient is a genuine nonperturbative symmetry of type-II string theory.

In section 3 we turn our analysis to type-I superstring theory and its
associated type-II orientifolds. The type-I theories were discussed in
\cite{witten,horava,bgh}, while the K-theoretic description of orbifolds
was examined in \cite{garcia} and an analysis of various orientifold theories
in \cite{gukov,hori,bgh}. Here we examine the weak Bott homotopy sequence
associated with real K-theory through a careful analysis of real spinor
representations, which now takes us naturally through a chain of type-II
orientifolds in various dimensions. From this analysis we are able to
completely classify the vacuum manifolds for all of these orientifold theories,
and from the ABS isomorphism we construct the explicit forms of the classical
tachyon scalar fields in each case. This in turn identifies the field content
on the brane worldvolumes which are relevant to the bound state constructions
in these theories. We show how the Bott periodicity theorem in these cases
naturally encodes information about the diversities of brane constructions in
various dimensions, and again compare with the bound state constructions of
\cite{senbps,seninst}. Generally, the K-theoretic description of type-I
superstring theory is only valid at weak string coupling, but $S$-duality can
still be used in a heuristic way to say something about the perturbative
spectrum of heterotic strings \cite{witten}. We briefly describe the dual
models related to each of these type-II orientifolds, and as an example we
discuss, using equivariant K-theory, some relationships to conventional
orbifolds of the type-II theories as have been described in
\cite{sendescent,bg}. These K-theoretical constructions could have important
ramifications for the structure of the moduli spaces of these theories, which
have been recently re-examined in \cite{motlbanks}. We also describe the
analogs of the $\klein$ transformations, pointing out the qualitative
differences with the corresponding projections of the type-II models.

In section 4 we present a more systematic analysis of the periodicity
properties of the orientifold models through a more thorough description of the
stable equivariant homotopy sequence and the ABS construction for Real
K-theory\cite{atiyahreal}. We show that the extended Bott periodicity here
implies an unexpected descent relation whereby a $p-2$-brane localized at a
singular plane of a type-II orientifold theory is realized as a
$\zed_2$-equivariant magnetic monopole in the worldvolume of a
$p$-$\overline{p}$ brane pair located over an orientifold plane of one higher
dimension. These structures illuminate the internal symmetries inherent in the
Real K-theory and also how these groups take into account the RR-charges
carried by the orientifold planes. We also illustrate how Real K-theory
encompasses all of the brane constructions in a unifying
description. Within this description, the fundamental brane constructions are
all determined as solitonic configurations associated with the four Hopf
fibrations. This determines string solitons in terms of magnetic monopoles in
the type-II theories, while in the type-I theories we obtain non-BPS branes as
kinks, BPS branes as $SU(2)$ instantons, and both BPS and non-BPS branes as
$spin(8)$ instantons (along with their equivariant versions in the case of
orbifold and orientifold models). The internal orientifold symmetries are then
determined by equivariant monopoles which also relate both supersymmetric and
non-supersymmetric D-branes.

\newsubsection{K-theory Conventions}

In the remainder of this section we shall introduce the relevant K-theoretical
conventions that will be used in this paper. For concise introductions to the
subject, see \cite{karoubi,spingeom,husemoller,atiyah}. Given a compact
topological space $X$, the Grothendieck group $K(X)$ is defined as the set of
equivalence classes of pairs of complex vector bundles $[(E,F)]=[E]-[F]$ over
$X$, where $(E,F)\equiv(E\oplus G,F\oplus G')$ for any pair of vector bundles
$(G,G')$
(which will correspond to brane-antibrane creation and annihilation with
respect to the supersymmetric vacuum). The group operation is induced by the
Whitney sum of vector bundles and the inverse of the class $[(E,F)]$ is
$[(F,E)]$. The map $X\to K(X)$ is a contravariant functor from the category of
compact topological spaces to the category of abelian groups. The inclusion
$i:{\rm pt}\hookrightarrow X$ and the projection $X\to\rm pt$ with respect to a
fixed basepoint of $X$ induce a split short exact sequence in cohomology which
leads to the decomposition
\beq
K(X)= \widetilde{K}(X)\oplus K({\rm pt})
\label{KXdecomp}\eeq
with $K({\rm pt})=\zed$. The subgroup $\widetilde{K}(X)=\ker i^*$ is the
reduced K-group of $X$ consisting of classes of pairs of vector bundles of
equal rank (as will be required by tadpole anomaly cancellation in type-I and
type-IIB superstring theory). We shall only work in K-theory with compact
support (this will correspond to the statement that a brane has finite
tension). This means that for each class $[(E,F)]$, there is a map $T:E\to F$
which is an isomorphism of vector bundles outside an open set $U\subset X$
whose closure $\overline U$ is compact ($U$ will represent the region of the
transverse space where the topological soliton charge is localized while $T$
will be the tachyon field of a given unstable configuration of branes). This
condition automatically implies that $E$ and $F$ have the same rank, and hence
we shall mostly deal with the reduced K-group $\widetilde{K}(X)$ (this means
that we always measure brane charges with respect to that of the supersymmetric
vacuum). The corresponding virtual bundle may then be represented as
\beq
\Bigl[(E\,,\,F)\Bigr]=\Bigl[(\ker T\,,\,{\rm coker}\,T)\Bigr]
\label{grothrep}\eeq
When $X$ is not compact, we define $K(X)=\widetilde{K}(X^+)$, where $X^+$ is
the one-point compactification of $X$. We shall denote the trivial bundle of
rank $k$ over $X$ by $I^k$.

The higher K-group $K^{-1}(X)$ is defined as the abelian group of equivalence
classes of pairs $[(E_\alpha,E_{\,\ids})]$, where $\alpha$ is an
automorphism of $E$ ($\id$ denotes the identity automorphism) and $E_\alpha$ is
the vector bundle over $S^1\times X$ with total space $[0,1]\times E$ modulo
the identification $(0,v)\equiv(1,\alpha(v))~~\forall v\in E$. With this
definition we have $K^{-1}(X)\subset\widetilde{K}(S^1\times X)$ and
$K^{-1}({\rm pt})=0$, so that $K^{-1}(X)=\widetilde{K}^{-1}(X)$. Similarly one
also defines higher degree K-groups $K^{-n}(X)$, with $K^0(X)=K(X)$.

We shall be frequently interested in the K-groups of the product of two spaces
$Y$ and $W$. For this, we introduce the reduced join $Y\vee W$ of $Y$ and $W$,
i.e. their disjoint union with a basepoint of each space identified, which can
be viewed as the closed subspace $Y\times{\rm pt}\cup{\rm pt}\times W$ of the
Cartesian product $Y\times W$. Let $Y\wedge W=Y\times W/Y\vee W$ be the smash
product of $Y$ and $W$. Note that when $W=S^1$, $Y\wedge S^1=\Sigma\,Y$ is the
reduced suspension of the topological space $Y$, so that
\beq
\widetilde{K}^{-n}(\Sigma\,Y)=\widetilde{K}^{-n-1}(Y)
\label{Ksusp}\eeq
The inclusion $Y\vee W\hookrightarrow Y\times W$ and the canonical projection
$Y\times W\to Y\wedge W$ induce a split short exact sequence in reduced
K-theory, leading to
\bea
\widetilde{K}^{-n}(Y\times W)&=&\widetilde{K}^{-n}(Y\vee
W)\oplus\widetilde{K}^{-n}(Y\wedge
W)\nn\\&=&\widetilde{K}^{-n}(Y)\oplus\widetilde{K}^{-n}(W)
\oplus\widetilde{K}^{-n}(Y\wedge W)
\label{Kiso}\eea
The tensor product of vector bundles induces a cup product which gives
$K(X)$ the structure of a $\zed_2$-graded ring. This yields the canonical
homomorphisms
\bea
K(Y)\otimes_\zeds K(W)&\longrightarrow&K(Y\times
W)\nn\\\widetilde{K}(Y)\otimes_\zeds\widetilde{K}(W)
&\longrightarrow&\widetilde{K}(Y\wedge W)
\label{kunneth}\eea
which are induced by the cup product and the canonical projection in
\eqn{Kiso}. When either $K(Y)$ or $K(W)$ is a free abelian group, the mappings
in \eqn{kunneth} are isomorphisms, leading to the usual cohomological K\"unneth
theorem.

\newsection{Descent Equations in Type-II Theories}

The basic tool in the K-theory correspondence for type-II theories is the ABS
construction for the complex K-groups of even dimensional spheres \cite{abs}.
Let $spin(m)$ be the spin group of dimension $2^{[\frac m2]}$ which is a double
cover of the isometry group $SO(m)$ of the sphere $S^{m-1}$. Let $\Delta_{2n+1}$
be the $spin(2n+1)$-module corresponding to the unique irreducible
representation of the complexified Clifford algebra $C_{2n+1}^c=\complex(2^n)$
of dimension $2^n$, and let $\Delta_{2n}^\pm$ denote the $spin(2n)$-modules
corresponding to the two $2^{n-1}$ dimensional irreducible representations of
$C_{2n}^c=\complex(2^{n-1})\oplus\complex(2^{n-1})$ (Generally, ${\bb F}(m)$
denotes the $\real$-algebra of $m\times m$ matrices with entries in the
field $\bb F$). Let $R[spin(m)]$ be the (complex) representation ring of
$spin(m)$, i.e. the Grothendieck group constructed from the abelian monoid
generated by the irreducible
representations, with respect to the direct sum and tensor product of
$spin(m)$-modules. Then the natural embedding $spin(2n)\hookrightarrow
spin(2n+1)$ induces the graded ring isomorphism \cite{abs}
\beq
R\Bigl[spin(2n)\Bigr]\,/\,R\Bigl[spin(2n+1)\Bigr]~\cong~\widetilde{K}(S^{2n})
\label{absisoK}\eeq
The groups in \eqn{absisoK} are isomorphic to $\zed$,
while they would vanish for odd-dimensional spheres. This leads to the usual
integer spectrum of alternating dimension D$p$-brane charges (with $p$ odd for
IIB and even for IIA branes).

\newsubsection{Bott Periodicity}

For type-II D-branes, the basic relation we shall study is the weak Bott
periodicity theorem for stable homotopy groups of unitary homogeneous spaces,
\beq
\dots~{\buildrel\approx\over\longrightarrow}
{}~\pi_{2k-1}\Bigl(U(N)\Bigr)~{\buildrel\approx
\over\longrightarrow}~\pi_{2k}\Bigl(U(2N)/[U(N)\times
U(N)]\Bigr)~{\buildrel\approx\over\longrightarrow}
{}~\pi_{2k+1}\Bigl(U(2N)\Bigr)~{\buildrel\approx\over
\longrightarrow}~\dots
\label{botthomotopy}\eeq
where $N=2^{k-1}$ and $2k=9-p$ is the codimension of a type-IIB $p$-brane
worldvolume ${\cal M}_{p+1}$ in the spacetime manifold $X$. Our first
observation will be that the $T$-duality mapping between the type-IIB and
type-IIA theories is precisely the natural isomorphism between homotopy groups
at each step in \eqn{botthomotopy}, showing how a $p$-brane of IIB is mapped
into a $p-1$-brane of IIA, and vice versa. In this subsection we shall start by
describing the details of the isomorphism at the level of the homotopy groups
\cite{bfn}.

Consider a $p$-brane in the type-IIB theory constructed as the tachyonic
soliton of a bound state of $N=2^{k-1}$ 9-brane $\overline{9}$-brane pairs
\cite{witten}. The rank $2^k$ spinor bundle of the $2k$-dimensional transverse
space has a natural grading ${\cal S}_N^+\oplus{\cal S}_N^-$ induced by the
chirality grading of the associated Clifford bundle, where ${\cal S}_N^\pm$ are
the chiral spinor bundles of rank $2^{k-1}$ which carry the irreducible
representation $\Delta_{2k}^\pm$ of the Clifford algebra of the transverse
space. The generators $\Gamma_i$ of
$\Delta_{2k}^+\oplus\Delta_{2k}^-$ act off-diagonally as
$\Gamma_i:\Delta_{2k}^\pm\to\Delta_{2k}^\mp$. The spinor bundles may be
extended over the whole spacetime $X$ \cite{witten}. The bundle ${\cal S}_N^+$
(resp. ${\cal
S}_N^-$) is then identified as the Chan-Paton bundle carried by the
9-branes (resp. $\overline{9}$-branes), so that these spinor bundles produce a
K-theory class
$[({\cal S}_N^+,{\cal S}_N^-)]=[{\cal S}_N^+]-[{\cal
S}_N^-]\in\widetilde{K}(X)$. The gauge symmetry on the spacetime filling
9-brane worldvolume is $U(N)\times U(N)$. The tachyon field lives in the
bifundamental ${\bf N}\otimes\overline{\bf N}$ representation of the
gauge group (as required by tadpole anomaly cancellation) and is a map
$T_N^{\rm(B)}:{\cal S}_N^+\to{\cal
S}_N^-$. It vanishes on the worldvolume ${\cal M}_{p+1}$ and approaches its
vacuum expectation value $T_N^{\rm(B)0}$ at infinity in $X$, where we assume
that the eigenvalues of $T_N^{\rm(B)0}$ all have the same modulus
\cite{witten,seninst}. It therefore breaks the 9-brane gauge symmetry to the
type-IIB vacuum manifold $U(N)\times U(N)/U(N)_{\rm diag}= U(N)$
(topologically) which represents the stable vortex configurations of the
tachyon field. The $p$-brane charge is determined by the winding number of the
tachyon field at infinity which generates the homotopy group
$\pi_{2k-1}(U(N))$. The explicit representation of the tachyonic configuration
is via Clifford multiplication by an element of the sphere bundle
\eqn{spherebundle} for $n=2k$,\footnote{Here and in the following it is always
implicitly understood that the tachyon field is multiplied by a convergence
factor which approaches 1 near ${\cal M}_{p+1}$ (so that the soliton is located
at $x^i=0$) and ensures that at $|x|\to\infty$, $T_N^{\rm(B)}(x)$ takes values
in the type-IIB vacuum manifold, as in \eqn{puregauge}.}
\beq
T_N^{\rm(B)}(x)=\sum_{i=1}^{2k}\Gamma_i\,x^i=\sum_{i=1}^{2k}
\pmatrix{0&\gamma_i^+\,x^i\cr\gamma_i^-\,x^i&0\cr}
\label{tachyonIIB}\eeq
where $x^i$ are local coordinates of the transverse space to
${\cal M}_{p+1}$ in $X$ and $\gamma_i^\pm$ are the generators of
$\Delta_{2k}^\pm$. The prescription described above embeds the K-group
$\widetilde{K}({\cal M}_{p+1})$ of the $p$-brane worldvolume into the spacetime
K-group $\widetilde{K}(X)$, such that the RR-charge takes values in the K-group
\beq
\widetilde{K}(S^{2k})=\pi_{2k-1}\Bigl(U(N)\Bigr)=\zed
\label{KS2k}\eeq
of the transverse space. The precise mapping is given via the cup product in
\eqn{kunneth}, leading to
\bea
\lambda_N\,:\,K({\cal M}_{p+1})\otimes_\zeds
K(S^{2k})&{\buildrel\approx\over\longrightarrow}&K(X)\nn\\
\Bigl[(E\,,\,F)\Bigr]&\mapsto&\lambda_N\left(\left[(E\otimes{\cal S}_N^+
\oplus F\otimes{\cal S}_N^-\,,\,E\otimes{\cal S}_N^-\oplus
F\otimes{\cal S}_N^+)\right]\right)\nn\\& &
\label{absmapK}\eea
where $[(E,F)]\in\widetilde{K}({\cal M}_{p+1})$ and we have used the fact that
$K(S^m)$ for any $m$ is a free abelian group.

There is a canonical mapping of this system onto a configuration in the
type-IIA theory representing a $p-1$-brane of codimension $2k+1$ in $X$,
constructed as the tachyonic vortex of a system of unstable 9-branes
\cite{horava}.
There are now $2N$ 9-branes (unstable $\overline{9}$-branes in the IIA theory
are indistinguishable from unstable 9-branes as they carry no conserved
charge), with gauge symmetry $U(2N)$, whose tachyon condensate $T_N^{\rm(A)0}$
has an equal number of positive and negative eigenvalues \cite{horava}. The
9-brane gauge symmetry is therefore broken to the type-IIA vacuum manifold
$U(2N)/[U(N)\times U(N)]$ representing the stable soliton configurations of the
tachyon field $T^{\rm(A)}_N$ which lives in the adjoint representation of the
$U(2N)$ gauge group and generates the homotopy group
$\pi_{2k}(U(2N)/[U(N)\times U(N)])$. Promoting one of the coordinates $x^{p+1}$
of ${\cal M}_{p+1}$ to the
transverse space, we identify the $U(2N)$ Chan-Paton bundle carried by the
9-branes as the irreducible rank $2^k$ spinor bundle ${\cal S}_N$ of the $2k+1$
dimensional transverse space to the new $p-1$-brane worldvolume ${\cal M}_p$.
Then the tachyon field is a map $T^{\rm(A)}_N:{\cal S}_N\to{\cal S}_N$ which
produces a higher degree K-theory class $[({\cal S}_{\tau_N},{\cal
S}_{\,\ids})]\in K^{-1}(X)$, where
\beq
\tau_N=-\exp\pi i\,T^{\rm(A)}_N
\label{tauN}\eeq
acts by the natural adjoint action on ${\cal S}_N$. The natural embedding
$spin(2k)\hookrightarrow spin(2k+1)$ identifies the corresponding $2k+1$
generators of $\Delta_{2k+1}$ as the $\Gamma_i$ in \eqn{tachyonIIB} along
with the chirality matrix
\beq
\Gamma_{2k+1}=(-i)^k\,\Gamma_1\cdots\Gamma_{2k}=(\sigma_3)^{\otimes k}
\label{GammaIIA}\eeq
where $\sigma_i$ will always denote the standard Pauli spin matrices. This
gives a natural map $T_N^{\rm(B)}(x)\mapsto
T_N^{\rm(A)}(x,x^{p+1})=\sum_i\Gamma_i\,x^i+\Gamma_{2k+1}\,x^{p+1}$ on
$\pi_{2k-1}(U(N))\to\pi_{2k}(U(2N)/[U(N)\times U(N)])$, i.e.
\beq
T^{\rm(A)}_N(x,x^{p+1})=T_N^{\rm(B)}(x)+\pmatrix{x^{p+1}\,I_N&0
\cr0&-x^{p+1}\,I_N\cr}
\label{tachyonIIA}\eeq
where $T_N^{\rm(B)}(x)$ is the IIB tachyon field \eqn{tachyonIIB} and $I_N$ is
the $N\times N$ identity matrix. The map \eqn{tachyonIIA} generates the higher
degree K-group
\beq
K^{-1}(S^{2k+1})=\pi_{2k}\Bigl(U(2N)/[U(N)\times U(N)]\Bigr)=\zed
\label{KS2k1}\eeq
of the transverse space which labels the induced $p-1$-brane RR charge. The
construction above defines an embedding $\widetilde{K}({\cal M}_p)\to
K^{-1}(X)$, again via the cup product \eqn{kunneth} giving
\bea
\widehat{\lambda}_N\,:\,K({\cal M}_{p})\otimes_\zeds
K^{-1}(S^{2k+1})&{\buildrel\approx\over\longrightarrow}&
K^{-1}(X)\nn\\\Bigl[(E\,,\,F)\Bigr]&\mapsto&\widehat{\lambda}_N
\left(\left[\Bigl((E\otimes{\cal S}_N)_{\ids\otimes\tau_N}
\,,\,(F\otimes{\cal S}_N)_{\ids\otimes\tau_N}\Bigr)\right]\right)\nn\\& &
\label{absmapK1}\eea
for $[(E,F)]\in\widetilde{K}({\cal M}_p)$. The transformation described here is
actually nothing but the $T$-duality mapping
between the type-IIB and type-IIA theories. Starting with the type-IIB theory,
a $T$-duality transformation along one of the longitudinal directions of the
$p$-brane transforms it into a transverse space direction and maps the
$p$-brane onto a $p-1$-brane. It acts on the K-theory classes of $X$ by mapping
the 9-brane-antibrane pairs onto $N$ 8-brane-antibrane pairs, of codimension 1
in $X$, which can each be represented as the tachyonic kink of an unstable
9-brane in the type-IIA theory \cite{horava}. This is represented by the
diagonal matrix in \eqn{tachyonIIA}. The $p-1$-brane itself is then represented
as the codimension $2k$ tachyonic soliton of the bound state of the 8-brane
$\overline{8}$-brane pairs connected together by the same tachyon field
$T_N^{\rm(B)}(x)$ as in the IIB case.

Finally, we come to the second isomorphism in \eqn{botthomotopy}, which gives a
natural map from the type-IIA system above to the type-IIB theory describing a
$p-2$-brane, with worldvolume ${\cal M}_{p-1}$ obtained by promoting the
coordinate $x^p$ of ${\cal M}_p$ to the transverse space, which is described in
terms of the bound state of $2N$ 9-brane $\overline{9}$-brane pairs. The spinor
representation of $SO(2k)$ may be mapped into that of $SO(2k+2)$ by defining
$2^{k+1}$ dimensional generators $\widehat{\Gamma}_i$ of
$\Delta_{2k+2}^+\oplus\Delta_{2k+2}^-$ via
\beq
\widehat{\Gamma}_i=\sigma_3\otimes\Gamma_i
{}~~~~~~,~~~~~~\widehat{\Gamma}_{2k+1,2k+2}=\sigma_{1,2}\otimes I_{2N}
\label{hatgens}\eeq
The new tachyon configuration whose vortex core corresponds to the $p-2$-brane
worldvolume and which generates $\widetilde{K}(S^{2k+2})=\pi_{2k+1}(U(2N))$ is
thus
$T_{2N}^{\rm(B)}(x,x^p,x^{p+1})=\sum_i\widehat{\Gamma}_i\,x^i
+\widehat{\Gamma}_{2k+1}\,x^p+\widehat{\Gamma}_{2k+2}\,x^{p+1}$, i.e.
\beq
T_{2N}^{\rm(B)}(x,x^p,x^{p+1})=\pmatrix{T_N^{\rm(B)}(x)&0\cr0
&-T_N^{\rm(B)}(x)\cr}
+\pmatrix{0&\left(x^p+i\,x^{p+1}\right)I_{2N}\cr
\left(x^p-i\,x^{p+1}\right)I_{2N}&0\cr}
\label{tachyonN+1}\eeq
Again the form of \eqn{tachyonN+1} is naturally explained by $T$-duality.
Applying a $T$-duality transformation to the IIA system above maps the 8-branes
and $\overline{8}$-branes to $N$ 7-branes and $\overline{7}$-branes, which are
each of codimension 2 in $X$ and whose tachyonic vortex representation in terms
of 9-brane-antibrane pairs is given by the off-diagonal block matrix in
\eqn{tachyonN+1} \cite{horava}. The $p-2$-brane is represented as the tachyonic
soliton in the 7-brane worldvolume. In this case, the unstable 9-branes of the
IIA-theory are mapped into unstable 8-branes of the IIB-theory, each of which
gives rise to a tachyonic kink representing a 7-brane or $\overline{7}$-brane.
The result is the addition of an extra set of $N$ 7-$\overline{7}$ pairs
required for K-theoretic stability \cite{witten,horava} of the bound state
construction, so that the full system of 7-branes are connected together by the
tachyon field given by the block diagonal matrix in \eqn{tachyonN+1}. In this
stabilizing operation, the extra set of 7-branes does not interact via the
tachyon field with the original set of 7-branes.

The intermediate configuration \eqn{tachyonIIA} can also be regarded
as a $p-1$-brane of the type-IIB theory, obtained as the codimension 1
tachyonic kink of a $p$-$\overline p$ brane pair. However, the tachyon field
$T_N^{\rm(A)}$ has winding number 0 in codimension $2k$ and so this
configuration maps to the trivial (identity) element of $\widetilde{K}(X)$.
This is simply the K-theoretic statement that the $p-1$-brane is an unstable
non-BPS configuration of type-IIB superstring theory. It becomes stable upon
another tachyon condensation in codimension 1, giving
\eqn{tachyonN+1}.\footnote{Notice that the above construction (and the others
to follow) applies equally well starting in the type-IIA theory with a brane
configuration of codimension $2k+1$ in $X$, by decomposing the spin group
$spin(2k+1)\supset spin(2k)$ and applying the construction to the $2k$
dimensional subspace.} The map between the tachyon fields \eqn{tachyonIIB} and
\eqn{tachyonN+1} (relating $p$-branes and $p-2$-branes in the IIB theory) is a
typical example
of the strong Bott periodicity isomorphism in complex K-theory \cite{karoubi}
\beq
\widetilde{K}^{-n}(X)=\widetilde{K}^{-n-2}(X)
\label{bottKn}\eeq
which, according to \eqn{absisoK} and the above construction, can be described
by the ABS map \cite{abs}
\bea
\left[({\cal S}_N^+\,,\,{\cal S}_N^-)\right]&\mapsto&\left[\left({\cal
S}_N^+\otimes({\cal S}_1^+\oplus{\cal S}_1^-)\,,\,{\cal S}_N^-\otimes({\cal
S}_1^+\oplus{\cal S}_1^-)\right)\right]\nn\\T_N&\mapsto&\pmatrix{T_N\otimes
I_2&-I_{2N}\otimes T_1^\dagger\cr I_{2N}\otimes T_1&T_N^\dagger\otimes
I_2\cr}
\label{Bottspinmap}\eea
The block diagonal matrix in \eqn{tachyonN+1} corresponds to the representation
of $p$-branes and $\overline{p}$-branes each as bound states of $N=2^{k-1}$
spacetime-filling 9-brane-antibrane pairs, while the off-diagonal block matrix
in \eqn{tachyonN+1} corresponds to the tachyonic vortex configuration of a
$p-2$-brane constructed as the bound state of the $p$-$\overline p$ pair in
codimension 2. Thus the transformation from \eqn{tachyonIIB} to
\eqn{tachyonN+1} represents the process of tachyon condensation of the bound
state of a $p$-brane $\overline{p}$-brane pair into a $p-2$-brane, regarded
as the Bott periodicity map \eqn{Bottspinmap} on the spacetime K-theory group
$\widetilde{K}(X)\to\widetilde{K}(X)$.

The mod 2 periodicity \eqn{bottKn} follows from the periodicity
$C_{l+2}^c=C_l^c\otimes\complex(2)$ of complexified Clifford algebras
\cite{spingeom}, and the isomorphism \eqn{Bottspinmap} comes from the cup
product on the K-groups. Taking $W=S^2$ and $Y$ to be the $n$-th reduced
suspension of $X$ in \eqn{kunneth} gives
$\widetilde{K}(\Sigma^nX)\otimes_\zeds\widetilde{K}(S^2)
=\widetilde{K}(\Sigma^nX\wedge S^2)$, which using \eqn{Ksusp} yields the
isomorphism
\beq
\alpha\,:\,\widetilde{K}^{-n}(X)\otimes_\zeds\widetilde{K}(S^2)~
{\buildrel\approx\over\longrightarrow}~\widetilde{K}^{-n-2}(X)
\label{alphaiso}\eeq
The generator $[{\cal N}_\complexs]-[I^1]$ of $\widetilde{K}(S^2)=\zed$ may be
described by taking ${\cal N}_\complexs$ to be the canonical line
bundle over $\complex P^1$, which is associated with the Hopf fibration $S^3\to
S^2$. Then the isomorphism \eqn{bottKn} is given by the mapping
\beq
\Bigl[(E\,,\,F)\Bigr]~\mapsto~\alpha\left(\Bigl[(E\otimes{\cal
N}_\complexs\,,\,F\otimes{\cal N}_\complexs)\Bigr]\right)
\label{altBottmap}\eeq
for $[(E,F)]\in\widetilde{K}^{-n}(X)$. This construction shows that the
codimension 2 tachyonic soliton in the worldvolume of a $p$-$\overline{p}$
brane pair which represents a type-II $p-2$-brane may be identified with the
usual
Dirac monopole associated with the complex Hopf bundle ${\cal N}_\complexs$
\cite{trautman}. This K-theoretic fact agrees with the construction of
\cite{seninst} of a vortex-type solution on the membrane-antimembrane pair in
type-IIA string theory, whereby the asymptotic field configurations
\eqn{puregauge} on the membrane worldvolume resemble exactly those of a
magnetic monopole configuration with charges living in $\pi_1(U(1))=\zed$.

\newsubsection{$T$-duality Transformations}

To formalize the $T$-duality transformation of the spacetime K-groups, we
compactify one of the $p$-brane worldvolume directions on a circle $S^1$, so
that the spacetime manifold is now the product space $X=Y\times S^1$. To study
the K-theory of this compactification, we use \eqn{Ksusp} and the product
formula \eqn{Kiso} with $W=S^1$ to get
\bea
\widetilde{K}(Y\times
S^1)&=&K^{-1}(Y)\oplus\widetilde{K}(Y)\label{wKXciso}\\K^{-1}(Y\times
S^1)&=&\left(\widetilde{K}(Y)\oplus\zed\right)\oplus K^{-1}(Y)
\label{K1Xciso}\eea
where we have used Bott periodicity. The splitting of the K-groups here can be
understood by noting that the natural embedding $i:Y\hookrightarrow Y\times
S^1$ induces a projection on K-theory $i^*:\widetilde{K}^{-n}(Y\times
S^1)\to\widetilde{K}^{-n}(Y)$ such that $\ker
i^*=\widetilde{K}^{-n-1}(Y)\oplus\widetilde{K}^{-n}(S^1)$. It follows that the
group $\widetilde{K}^{-n}(Y)$ labels the corresponding Kaluza-Klein modes that
arise from the compactification of the spacetime on $S^1$. The other factor
$\widetilde{K}^{-n-1}(Y)\oplus\widetilde{K}^{-n}(S^1)$ represents the unwrapped
modes of the D-brane configurations. That this interpretation is indeed precise
can be proven by taking $W$ to be the topological space consisting of a single
point in \eqn{Kiso}, which gives
\beq
\widetilde{K}^{-n}(Y\times{\rm pt})=\widetilde{K}^{-n}(Y)
\label{Kptiso}\eeq
where we have used $\widetilde{K}^{-n}({\rm pt})=0$. Eq. \eqn{Kptiso} shows
that as the compactified direction is shrunk to a point, only the Kaluza-Klein
modes contribute to the D-brane charges which are now determined by the
K-theory classes of the uncompactified nine dimensional space $Y$.

The extra integer subgroup in the IIA case \eqn{K1Xciso} relative to the IIB
case \eqn{wKXciso} labels the large gauge transformations of the tachyon field
\eqn{tachyonIIA} around the compactified direction $x^{p+1}\in S^1$. The
$T_N^{\rm(B)}$ part of the IIA tachyon field \eqn{tachyonIIA} generates the
$\widetilde{K}(Y)$ subgroup of \eqn{K1Xciso} representing the unwrapped brane
charges, and the map $T^{\rm(A)}_N\mapsto\tau_N$ in \eqn{tauN} used to define
an element of $K^{-1}(X)$ has kernel which is isomorphic to $\zed$. It is
invariant under the shift
\beq
T^{\rm(A)}_N\to T^{\rm(A)}_N+w\,\sigma_3\otimes I_N~~~~~~,~~~~~~w\in\zed
\label{TNgaugetransf}\eeq
corresponding to the windings of the tachyon field around the $S^1$. More
precisely, the IIB tachyon field $T_N^{\rm(B)}(x)$ is the transition function
on the overlap $S_+^{2k}\cap S_-^{2k}= S^{2k-1}$ (generating
$\pi_{2k-1}(U(N))$) that allows one to piece together topologically trivial
gauge fields on the contractible upper and lower hemispheres $S_\pm^{2k}\subset
S^{2k}$ to obtain a gauge field on $S^{2k}$ with non-trivial topological $U(N)$
charge. This configuration represents the unbroken part of the $U(2N)$ gauge
field carried by the unstable 9-branes of the IIA theory \cite{horava}, showing
how the large gauge transformations arise upon target space compactification.
On the other hand, the automorphism \eqn{tauN} generates $\pi_{2k+1}(U(2N))$
\cite{horava,bfn}, so that the effect of the large gauge transformations
disappears when going from the IIA theory to the IIB theory. Thus the
$T$-duality mapping in \eqn{wKXciso} and \eqn{K1Xciso} shows explicitly how
under $T$-duality a longitudinal brane coordinate is mapped onto a gauge field
configuration winding around the compactified direction, and moreover how
$T$-duality interchanges Kaluza-Klein modes and unwrapped D-brane
configurations. The isomorphism mod $\zed$ between \eqn{wKXciso} and
\eqn{K1Xciso} is therefore precisely the natural transformation between the IIB
and IIA tachyon generators, regarded as the Bott periodicity isomorphism on
$\widetilde{K}(Y\times S^1)\to K^{-1}(Y\times S^1)$.

At the level of {\it unreduced} K-theory, the relation \eqn{KXdecomp} along
with \eqn{wKXciso} and \eqn{K1Xciso} show that the $T$-duality mapping is
indeed a natural isomorphism of the spacetime K-groups.\footnote{The extra
charges which appear in (\ref{wKXciso},\ref{K1Xciso}) (see also
(\ref{KTm},\ref{K1Tm}) and section 3.2) may be attributed
to vacuum charges that can be removed by adding a copy of the relevant
compactification manifold at infinity \cite{bgh}. The $T$-duality isomorphism
is then generated at the level of the corresponding {\it relative} K-theory
groups. This correspondence comes from the usual identification of brane
charges relative to that of the supersymmetric vacuum state. However,
insofar as the descent mechanisms which implement the $T$-duality
relationship between D-branes at the level of the Bott periodicity sequence
are concerned, the correct decompositions are as above with the extra
charges attributed to additional winding modes of the tachyon field
at infinity. We therefore keep this identification to correctly follow
the descent equations. Afterwards, one should identify the charges at
the level of relative K-theory, as discussed in \cite{bgh}.} However, the
isomorphism deteriorates in the decompactification limit, whereby only the
unwrapped D-brane configurations contribute to the charge. This can be seen
through the suspension isomorphism \cite{karoubi}
\beq
K^{-n}(Y\times\real)= K^{-n-1}(Y)
\label{unrediso}\eeq
which illustrates the usual result that the $T$-dual equivalence between the
IIB and IIA theories only holds upon compactification down to nine dimensions.
Upon descending to lower dimensional D-branes (as in \eqn{tachyonN+1}), we
encounter higher dimensional type-II toroidal compactifications.
The appropriate generalization of (\ref{wKXciso},\ref{K1Xciso}) may be found
inductively from \eqn{Ksusp} and \eqn{Kiso} to be
\bea
\widetilde{K}(Y\times
T^m)&=&K^{-1}(Y)^{\oplus2^{m-1}}\oplus\widetilde{K}(Y)^{\oplus2^{m-1}}
\oplus\zed^{\oplus(2^{m-1}-1)}\label{KTm}\\K^{-1}(Y\times T^m)
&=&\left(\widetilde{K}(Y)^{\oplus2^{m-1}}\oplus\zed\right)
\oplus K^{-1}(Y)^{\oplus2^{m-1}}\oplus\zed^{\oplus(2^{m-1}-1)}
\label{K1Tm}\eea
The K-groups \eqn{KTm} and \eqn{K1Tm} are again isomorphic mod $\zed$. Now the
decompositions correctly incorporate the dimensionality $2^{m-1}$ of the spinor
representation of the $T$-duality group $O(m,m;\zed)$ \cite{hori}, and hence of
the fact that IIB (resp. IIA) D-branes belong to the chiral (resp. antichiral)
spinor representations of $SO(2m)$ which are interchanged under a $T$-duality
transformation on $T^m$ with $m$ odd. These factors arise from the total number
of possible wrappings (spin structures) around the cycles of $T^m$. The
$\widetilde{K}(Y)^{\oplus2^{m-1}}$ subgroup of \eqn{K1Tm} representing
unwrapped brane charges is generated by the
$T_N^{\rm(B)}\otimes(\sigma_3)^{\otimes(m-1)}$ part of the IIA tachyon
generator $T_{2^mN}^{\rm(A)}$, while the extra
integer subgroups in both the IIA and IIB cases come from the higher degree
winding numbers of the tachyon fields (see \eqn{tachyonN+1}).

\newsubsection{$(-1)^{F_{\rm L}}$ Transformations}

We shall now describe the relationships between branes that arise via
application of the Klein operator $(-1)^{F_{\rm L}}$, where $F_{\rm L}$ is the
left-moving spacetime fermion number operator. Quotienting by $(-1)^{F_{\rm
L}}$ maps type-IIB superstring theory into the type-IIA theory. The operator
$\klein$ acts on the field content of the string theory by changing the sign of
all spacetime fields in the RR sector, and therefore the RR charge of a BPS
D-brane changes sign and it gets mapped to its antibrane under $\klein$. The
induced map on $\widetilde{K}(X)$ is the involution defined
by $[(E,F)]\to[(F,E)]$ and therefore the brane configurations which survive the
$\klein$-projection are those whose K-theory class is even under this $\zed_2$
action. The action on K-theory thereby induces a map
\beq
\widetilde{K}(X)\longrightarrow K^{-1}_{\zeds_2}(X\times\real^{0,1})
\label{Kequivmap}\eeq
where in general $\real^{p,q}$ is the $p+q$ dimensional real space in which an
involution acts as a reflection of the last $q$ coordinates, and $K_{\zeds_2}$
denotes the equivariant K-functor for the discrete group $\zed_2$ \cite{segal}.
It acts as the conventional K-functor on the category of $\zed_2$-equivariant
bundles over the space $X\times S^1$, i.e. the complex vector bundles $E$ whose
fiber projection $E\to X\times S^1$ commutes with the action of $\zed_2$. The
equivariant K-theory used in \eqn{Kequivmap} is a simplified version of the
Hopkins K-groups $K_\pm(X)$ \cite{witten,gukov} whereby, since $\klein$
acts only on the spectrum of the string theory, the $\zed_2$ action on $X\times
S^1$ is simply taken as an orientation reversing symmetry of $S^1$, with no
further geometrical action on the spacetime $X$.

We shall first give an elementary calculation of the right-hand side of
\eqn{Kequivmap} which will also prove useful later on when we discuss
orientifolds. The basic theorem we shall use is the six term exact sequence of
equivariant K-theory \cite{segal}
\beq\new{\begin{array}{ccccc}
K_{\zeds_2}^{-1}(M,A)&\longrightarrow&K_{\zeds_2}^{-1}(M)
&\longrightarrow&K_{\zeds_2}^{-1}(A)\\& & & &\\
{\scriptstyle\partial^*}\uparrow& & & &\downarrow
{\scriptstyle\partial^*}\\& & & &\\K_{\zeds_2}(A)
&\longleftarrow&K_{\zeds_2}(M)&\longleftarrow&K_{\zeds_2}
(M,A)\end{array}}
\label{sixterm}\eeq
where $A$ is a closed $\zed_2$-subspace of a locally compact $\zed_2$-space
$M$, and the relative K-theory is defined by
$K_{\zeds_2}^{-n}(M,A)=\widetilde{K}_{\zeds_2}^{-n}(M/A)$ (when the quotient
space makes sense). The horizontal maps in \eqn{sixterm} are induced by the
canonical inclusion and projection, while the vertical ones come from the
boundary homomorphisms $\partial$. The advantage of using this exact sequence
is that one may take $A$ to be the fixed point set of the group action on $M$,
such
that the quotient space $M/A$ has a free group action on it and its equivariant
cohomology can be computed as the ordinary cohomology of its quotient by
$\zed_2$.

In the present case, we take $M=X\times\real^{0,1}$ and $A=X\times\{0\}$. Then
$M/A$ is homotopic to two copies of $X\times\real$ which are exchanged by the
involution. Since the $\zed_2$ action on $M/A$ is free, the equivariant
K-groups may be computed by using the homotopy invariance of the K-functor and
the suspension isomorphism \eqn{unrediso} to get
\beq
K_{\zeds_2}^{-n}\Bigl((X\times\real)\amalg(X\times\real)\Bigr)
=K^{-n}(X\times\real)=K^{-n-1}(X)
\label{KMA}\eeq
On $A$ the $\zed_2$ action is trivial, so that
\beq
K_{\zeds_2}^{-n}(X\times\{0\})=K^{-n}(X\times{\rm pt})\otimes R[\zed_2]
\label{KA}\eeq
where
\beq
R[\zed_2]=\zed\oplus\zed
\label{repringZ2}\eeq
is the representation ring of the cyclic group $\zed_2$ \cite{segal}. Finally,
since in this
case $A$ is an equivariant retract of $M$, we have
$\ker\partial^*=K_{\zeds_2}^{-n}(A)$ and so the horizontal exact sequences in
\eqn{sixterm} split. In this way we arrive at
\beq
\widetilde{K}_{\zeds_2}^{-n}(X\times\real^{0,1})=\left(\widetilde{K}^{-n}(X)
\otimes R[\zed_2]\right)\oplus\widetilde{K}^{-n-1}(X)
\label{KZ21}\eeq
where we have used \eqn{Kptiso}. The doubling of the group
$\widetilde{K}^{-n}(X)$ in \eqn{KZ21} from its
product with the representation ring \eqn{repringZ2} just indicates that each
brane has a mirror image coming from the equivalence relation generated by
$\klein$. According to the above derivation it comes from the trivial part of
the $\zed_2$ action and as such represents the untwisted brane charges. The
other group $\widetilde{K}^{-n-1}(X)$ comes from the free part of the $\zed_2$
action and represents the twisted sector.

To construct explicitly the K-group mapping \eqn{Kequivmap} onto the type-IIA
theory, we need to act on type-IIB brane configurations (K-group classes) which
are invariant under the action of $\klein$. Such objects are the
$p$-$\overline{p}$ pairs corresponding to stable D$p$-brane configurations (on
which quotienting by $\klein$ makes sense). To describe this pair as an element
of the spacetime K-group $\widetilde{K}(X)$, we represent the $p$-brane as the
tachyonic vortex formed by $N=2^{k-1}$ 9-$\overline{9}$ brane pairs, which
define the K-theory class $[({\cal S}_N^+,{\cal S}_N^-)]\in\widetilde{K}(X)$
with tachyon generators $T_N^{{\rm(B)}\pm}$ acting on the chiral and antichiral
spinor bundles ${\cal S}_N^\pm$, as defined in \eqn{tachyonIIB}. Likewise, we
represent the $\overline{p}$-brane in terms of the bound state of another set
of $N$ 9-$\overline{9}$ brane pairs which define the spacetime K-theory class
$[(\overline{\cal S}_N^-,\overline{\cal S}_N^+)]$ with tachyon generators
$\overline{T}_N^{{\rm(B)}\pm}$, where $\overline{\cal S}_N^\pm$ are the
conjugate spinor bundles defined by complex-conjugating the
transition functions of ${\cal S}_N^\mp$ (note the change in sign of the RR
charge of the $\overline{p}$-brane relative to the $p$-brane). Of course,
$\overline{\cal S}_N^\pm={\cal S}_N^\pm$, but we shall keep the two sets of
bundles arbitrary to represent the pairing of a brane with its mirror antibrane
in \eqn{KZ21}. The $p$-$\overline{p}$ brane pair therefore determines the
K-theory class
\beq
\left[({\cal S}_N^+\,,\,{\cal S}_N^-)\right]+\left[(\overline{\cal
S}_N^-\,,\,\overline{\cal S}_N^+)\right]=\left[({\cal
S}_N^+\oplus\overline{\cal S}_N^-\,,\,{\cal S}_N^-\oplus\overline{\cal
S}_N^+)\right]\in\widetilde{K}(X)
\label{pbarpclass}\eeq
and tachyon field $T_N^{\rm(B)}\oplus(\overline{T}_N^{\rm(B)})^\dagger$. More
precisely, the class \eqn{pbarpclass} is determined by a pair $(E,F)$ of
Chan-Paton bundles each transforming under $SO(2k)$ rotations in the indicated
spinor representation $\Delta_{2k}^+\oplus\Delta_{2k}^-$ in their fibers. The
operator $\klein$ acts on the category of vector bundles by interchanging the
$p$-brane and $\overline{p}$-brane, i.e. ${\cal
S}_N^\pm\leftrightarrow\overline{\cal S}_N^\pm$, and on K-theory classes by
interchanging spacetime filling 9-branes and $\overline9$-branes, i.e.
$[(E,F)]\leftrightarrow[(F,E)]$ and
$T_N^{{\rm(B)}}\leftrightarrow(\overline{T}_N^{{\rm(B)}})^\dagger$. The K-group
elements of the form \eqn{pbarpclass} are therefore even under the action of
the Klein operator and correspond to the elements of the first direct summand
in \eqn{KZ21} for $n=1$ which survive the projection. The other direct summand
$\widetilde{K}(X)$ represents the IIB brane configurations which are projected
out of the spectrum by $\klein$.

The equivalence relation generated by the $\klein$ involution identifies the
classes $[{\cal S}_N^+\oplus\overline{\cal S}_N^-]\equiv[{\cal
S}_N^-\oplus\overline{\cal S}_N^+]$ when embedded as elements of the
equivariant K-group $K_{\zeds_2}^{-1}(X\times\real^{0,1})$. It follows that
there is a well-defined mapping from the subset of classes \eqn{pbarpclass} to
$K^{-1}(X)$ given by the projection
\beq
\left[({\cal S}_N^+\oplus\overline{\cal S}_N^-\,,\,{\cal
S}_N^-\oplus\overline{\cal S}_N^+)\right]~\mapsto~\left[\left(({\cal
S}_N^+\oplus\overline{\cal S}_N^-)_{-\exp\pi i\,\tilde T_N^{\rm(A)}}~,~({\cal
S}_N^+\oplus\overline{\cal S}_N^-)_{\ids\oplus\ids}\right)\right]\in K^{-1}(X)
\label{proj}\eeq
The canonical choice of IIA tachyon field $\tilde T_N^{\rm(A)}$ acting on
${\cal S}_N^+\oplus\overline{\cal S}_N^-$ is
$T_N^{{\rm(B)}+}\oplus\overline{T}_N^{{\rm(B)}-}$ (which is a well-defined
operator because of the equivalence relation). However, this tachyon
configuration has winding number 0 in codimension $2k+1$ and as such it is
trivial in $K^{-1}(X)$. This simply represents the fact that the $p$-$\overline
p$ brane pair of the IIB theory is mapped by the $(-1)^{F_{\rm L}}$-projection
to an unstable $p$-brane configuration in the IIA theory. To obtain a
non-trivial K-theory class, we must let the tachyon condense in codimension 1,
leading to the appropriate IIA tachyon field
\beq
\tilde
T_N^{\rm(A)}=T_N^{{\rm(B)}+}\oplus\overline{T}_N^{{\rm(B)}-}
+T^{(1)}\,\sigma_3\otimes I_N
\label{tildetachyonIIA}\eeq
where $T^{(1)}=x^{p+1}$ is the codimension 1 tachyon field. With
\eqn{tildetachyonIIA}, the class \eqn{proj} defines a non-trivial element of
$K^{-1}(X)$ in \eqn{KZ21}. This element is the representation of a stable
$p-1$-brane configuration as the tachyonic soliton of a collection of $2^{k-1}$
9-branes, used for the bound state construction of the original IIB $p$-brane,
and $2^{k-1}$ $\overline9$-branes, used to build the $\overline p$-brane.
Because of the equivalence relation in $K_{\zeds_2}^{-1}(X\times\real^{0,1})$,
the $\overline9$-branes are identified in the IIA theory as 9-branes, and hence
we obtain the usual K-theoretic representation of the $p-1$-brane in terms of
$2^k$ unstable 9-branes of type-IIA string theory.

The ``naive'' choice of tachyon field in \eqn{tildetachyonIIA} (represented by
the first term) suggests that one should consider it as a soliton configuration
of the IIB theory. To do so, we need a well-defined projection onto
$\widetilde{K}(X)$ . Since the ``naive'' tachyon field represents the identity
class of $K^{-1}(X)$, it is trivially invariant under an additional
$\klein$-projection (as it represents an unstable non-BPS configuration, it
carries no RR charge, and so is unaffected by the action of $\klein$).
Generally, the brane configurations which survive the modding out of the
type-II theory $m$ times by $\klein$ are represented by the K-group
\beq
\widetilde{K}_{\zeds_2}^{-n}(X\times\real^{0,m})=
\left(\widetilde{K}^{-n}(X)\otimes R[\zed_2]\right)\oplus
\widetilde{K}^{-n-m}(X)
\label{kleinm}\eeq
whose decomposition is found as before. For the case at hand, the relevant
group \eqn{kleinm} is obtained for $n=m=2$, so that the $\klein$-invariant
states reside in the first summand and the configurations which are projected
out are the same as those in \eqn{KZ21} for $n=1$. The equivalence relation
generated by the $\klein$ involution now gives $[({\cal S}_N^+,\overline{\cal
S}_N^-)]\equiv[({\cal S}_N^-,\overline{\cal S}_N^+)]$ as elements of the
equivariant cohomology $\widetilde{K}_{\zeds_2}(X\times\real^{0,2})$. The
well-defined projection onto the type-IIB theory is now represented by the
mapping of classes
\beq
\left[{\cal S}_N^+\oplus\overline{\cal S}_N^-\right]~\mapsto~\left[({\cal
S}_N^+\,,\,\overline{\cal S}_N^-)\right]\in\widetilde{K}(X)
\label{proj2}\eeq
and the corresponding tachyon field
\beq
\tilde T_N^{\rm(B)}=T_N^{{\rm(B)}+}\oplus\overline{T}_N^{{\rm(B)}-}
\label{tildetachyonIIB}\eeq
This class of $\widetilde{K}(X)$ represents a stable BPS D$p$-brane constructed
out of the bound state of $2^{k-1}$ 9-$\overline9$ brane pairs.

It is instructive to compare the K-theory construction above to the boundary
state formalism for the D-branes \cite{senbps}. A supersymmetric D$p$-brane
boundary state of the type-IIB theory is the sum of contributions from the
Neveu-Schwarz (NS) and Ramond sectors which is invariant under the appropriate
GSO projection,
\beq
|{\rm D}p\rangle=\mbox{$\frac12$}\Bigl(|Bp,+\rangle_{\rm NS}-|Bp,-\rangle_{\rm
NS}\Bigr)\pm\,\mbox{$\frac12$}\Bigl(|Bp,+\rangle_{\rm R}+|Bp,-\rangle_{\rm
R}\Bigr)
\label{Dpstate}\eeq
where the $\pm$ label the different spin structures, and the relative sign
between the NS-NS and RR contributions distinguishes a brane from its
antibrane. The RR contribution flips sign under the action of $\klein$, while
the NS-NS part is invariant. In order to describe the projection of the string
Hilbert space onto those states which are even under $\klein$, the natural
procedure is to cancel the odd RR part of the D$p$-brane boundary
state. This may be achieved by the superposition of a $p$-brane
with a $\overline{p}$-brane which is described by the boundary state
\beq
|\widetilde{Bp}\rangle=|{\rm D}p\rangle+|{\rm
D}\overline{p}\rangle=|Bp,+\rangle_{\rm NS}-|Bp,-\rangle_{\rm NS}
\label{Bpstate}\eeq
This state describes an unstable $p$-brane configuration of the type-IIA theory
\cite{horava} and is the configuration considered in \cite{sendescent}. On the
other hand, one can analyse the twisted and untwisted sectors of the string
Hilbert space and show that the result of quotienting by $\klein$ projects onto
the NS-NS parts of all IIB $p$-brane boundary states, as in \eqn{Bpstate}
\cite{daspark}. In this latter case, we find that the transformation on the
spacetime-filling 9-branes leaves an equal number of (identical) 9-branes and
$\overline{9}$-branes in the type-IIA theory. That this is also the
configuration that comes from quotienting the $p$-$\overline{p}$ system is a
natural consequence of the K-theoretic approach above. Namely, the final
configuration of $2^k$ unstable 9-branes of the type-IIA theory comes from
combining the $2^{k-1}$ 9-branes used in the bound state construction of the
$p$-brane and the $2^{k-1}$ $\overline{9}$-branes of the mirror
$\overline{p}$-brane, as is implicit in the analysis of \cite{sendescent}. This
is in contrast to the naive expectation that the $\klein$ projection would just
eliminate all $\overline{9}$-brane contributions. Moreover, the K-theory
derivation above shows why the twisted sectors of the $\klein$ projection do
not contribute to the RR-charge \cite{daspark}, in that the $K^{-1}(X)$
subgroup of \eqn{KZ21} comes from the trivial part of the $\zed_2$ action on
$X\times S^1$.

\newsubsection{Summary}

The relations among BPS and non-BPS D-branes in type-II superstring theory may
be succinctly summarized in the following diagram representing the K-theory
analog of fig. 1:
\beq\new{\begin{array}{rrcllll}
K(X)&{\buildrel\klein\over\longrightarrow}&
K_{\zeds_2}^{-1}(X\times\real^{0,1})&{\buildrel\klein\over\longrightarrow}
&K_{\zeds_2}^{-2}(X\times\real^{0,2})&{\buildrel{\Pi_1}\over\longrightarrow}&
K(X)\\~~&~~&~~&~~&~~&~~&\\{\scriptstyle\beta}\downarrow&~~&
{\scriptstyle\beta\circ\Pi_1}\downarrow&\!\swarrow&
\!\!\!\!\!\!\!\!\!\!\!\!\!{\scriptstyle K
(Y\times S^1)= K^{-1}(Y\times S^1)}&~~&\\~~&~~&~~&~~&~~
\\K^{-1}(X)&~~&K^{-1}(X)&~~&~~\\~~&~~&~~&~~&~~\\
{\scriptstyle\beta}\downarrow&\swarrow&\!\!\!\!\!
{\scriptstyle K(Y\times T^2)= K^{-1}(Y\times T^2)}&~~&~~
\\~~&~~&~~&~~&~~\\K(X)~=~K^{-2}(X)&~~&~~&~~&~~\end{array}}
\label{complexKdiag}\eeq
where $\beta$ is the Bott periodicity isomorphism representing tachyon
condensation in codimension 1, and $\Pi_1$ is the projection onto the first
direct summand of \eqn{kleinm}. The map $\beta^2$ realizes a type-II
$p-2$-brane as a Dirac magnetic monopole in the worldvolume of a
$p$-$\overline{p}$ brane pair. The $T$-duality transformations and the $\klein$
projections are formally identical, in that each one comes from mappings
between
subgroups of K-group decompositions in (\ref{wKXciso},\ref{K1Xciso}) and
\eqn{kleinm}, respectively (compare also \eqn{tachyonIIA} and
\eqn{tildetachyonIIA}). This shows how the K-theory description presents a
unified description of the D-brane descent relations and provides further
insight into the symmetry transformations between the IIA and IIB theories.

\newsection{Descent Equations in Type-I Theories}

Type-I superstring theory may be defined as the quotient of the type-IIB theory
by the twist operator $\Omega$ which reverses the orientation of the
fundamental string worldsheet. It acts on Chan-Paton bundles by conjugation and
therefore the D-brane configurations which survive the $\Omega$-projection are
those whose gauge bundles are real, i.e. have structure group $O(N)$. The
corresponding K-theory classes now live in the real K-group $KO(X)$
\cite{witten}. Many of the transformations described before based on the
structure of Clifford algebras carry through in the same way, except that one
must now taken into account the reality properties of the spinor modules
\cite{spingeom,husemoller}. The only real spinor modules (or, more precisely,
the only ones which are complexifications of real representations) are
$\Delta_{8k\pm1},\Delta_{8k}^\pm$, while the representations
$\Delta_{8k+3},\Delta_{8k+4}^\pm,\Delta_{8k+5}$ are the restrictions of
quaternionic Clifford modules. The remaining modules
$\Delta_{8k+2}^\pm,\Delta_{8k+6}^\pm$ are complex. This property modifies the
ABS isomorphism \eqn{absisoK} to \cite{abs}
\beq
\widetilde{KO}(S^{m})~\cong~RO\Bigl[spin(m)\Bigr]\,/\,RO\Bigl[spin(m+1)\Bigr]~
=~\left\{\new{\begin{array}{cll}\zed&~~&{\rm for}~~m\equiv0,4~
{\rm mod}~8\\\zed_2&~~&{\rm for}~~m\equiv1,2~{\rm mod}~8\\
0&~~&{\rm otherwise}\end{array}}\right.
\label{absisoKO}\eeq
where $RO[spin(m)]$ is the representation ring of real $spin(m)$-modules.

\newsubsection{Bott Periodicity and Type-II Orientifolds}

The weak Bott periodicity isomorphism for real K-theory has period 8 and may be
represented by the stable homotopy groups
\bea
\dots&{\buildrel\approx\over\longrightarrow}&\pi_{l-1}\Bigl(O(N)\Bigr)~
{\buildrel\approx\over\longrightarrow}~\pi_l\Bigl(O(2N)/[O(N)\times
O(N)]\Bigr)~{\buildrel\approx\over\longrightarrow}\nn\\&&\nn\\&{\buildrel
\approx\over\longrightarrow}&\pi_{l+1}\Bigl(U(2N)/O(2N)\Bigr)~
{\buildrel\approx\over\longrightarrow}~\pi_{l+2}\Bigl(Sp(2N)/U(2N)\Bigr)~
{\buildrel\approx\over\longrightarrow}\nn\\& &\nn\\&{\buildrel\approx\over
\longrightarrow}&\pi_{l+3}\Bigl(Sp(2N)\Bigr)~{
\buildrel\approx\over\longrightarrow}~\pi_
{l+4}\Bigl(Sp(4N)/[Sp(2N)\times
Sp(2N)]\Bigr)~{\buildrel\approx\over\longrightarrow}\nn\\&
&\nn\\&{\buildrel\approx\over\longrightarrow}&\pi_{l+5}\Bigl(U(8N)/Sp(4N)
\Bigr)~{\buildrel\approx\over\longrightarrow}~\pi_{l+6}\Bigl(O(16N)/U(8N)
\Bigr)~{\buildrel\approx\over\longrightarrow}~\pi_{l+7}\Bigl(O(16N)\Bigr)~
{\buildrel\approx\over\longrightarrow}~\dots\nn\\& &
\label{botthomKO}\eea
where $N=2^{[\frac l2]}$. For type-I branes the codimension $l=9-p$ may in
general be even or odd \cite{witten,senbps}. Thinking of the type-I theory as
the $\Omega$-projection of IIB superstring theory, a D-brane of codimension
$l=2k$ in $X$ comes from the bound state of $2^{k-1}$ type-IIB 9-$\overline{9}$
brane pairs with bundles $({\cal S}_{\frac N2}^+,{\cal S}_{\frac N2}^-)$
associated with the rank
$2^{k-1}$ spinor modules $\Delta_{2k}^\pm$. Since the 9-brane RR charge is
$\Omega$-invariant, we are left with $2^{k-1}$ 9-$\overline{9}$ brane pairs
plus their mirror images under the equivalence relation generated by $\Omega$.
Since the conjugate of ${\cal S}_{\frac N2}^\pm$ is ${\cal S}_{\frac N2}^\mp$,
the $2^{k-1}$ mirror 9-$\overline{9}$ pairs support the gauge bundles $({\cal
S}_{\frac N2}^-,{\cal S}_{\frac N2}^+)$ and so represent the corresponding
codimension $l$ antibrane. The total configuration of $2^k$
9-brane-antibrane pairs in the type-I theory therefore determines a K-theory
class $[(E,F)]$ where the fibers of the real gauge bundles $E$ and $F$ each
transform under $SO(2k)$ rotations as $\Delta_{2k}^+\oplus\Delta_{2k}^-$. As we
did in subsection 3.3, we may think of $[(E,F)]$ as a K-theory class $[({\cal
S}^+_{\frac N2}\oplus\overline{\cal S}_{\frac N2}^-,\overline{\cal S}^+_{\frac
N2}\oplus{\cal S}_{\frac N2}^-)]$ whereby branes are identified with their
antibranes under the equivalence relation generated by the given involution.
Similarly, when $l=2k+1$, the $p$-brane configuration is described in terms of
$2^k$
9-brane-antibrane pairs whose gauge bundles $(E,F)$ transform under the spinor
representation $\Delta_{2k+1}$ of the transverse rotation group.

In either case the tachyon field is given by
\beq
T_N^{\rm(I)}(x)=\sum_{i=1}^l\Gamma_i\,x^i
\label{TNI}\eeq
and it breaks the
$O(N)\times O(N)$ 9-brane gauge symmetry down to $O(N)_{\rm diag}$. The vacuum
manifold of the type-I theory is therefore homeomorphic to $O(N)$ and the
winding number of the tachyonic soliton configuration labels the induced
$p$-brane charge which lives in the real K-group
\beq
\widetilde{KO}(S^l)=\pi_{l-1}\Bigl(O(N)\Bigr)
\label{KOSgroups}\eeq
The integer-valued charges in \eqn{KOSgroups} and \eqn{absisoKO} correspond to
the usual stable type-I BPS D$p$-brane configurations for $p=1,5,9$ whose
type-IIB RR-charge is invariant under the $\Omega$-projection, i.e. whose
boundary states $|Bp,\pm\rangle_{\rm R}$ in \eqn{Dpstate} are even under
$\Omega$. The $\zed_2$-charges correspond to stable but non-BPS type-I
D$p$-brane configurations for $p=-1,0,7,8$. For $p=-1,7$ they may be
constructed as the $\Omega$-projection of the corresponding $p$-brane-antibrane
configuration of the type-IIB theory. For these values of $p$, $\Omega$
exchanges IIB $p$-branes with $\overline{p}$-branes, i.e.
$\Omega\,|Bp,\pm\rangle_{\rm R}=-|Bp,\pm\rangle_{\rm R}$, so that the
combination \eqn{Bpstate} is $\Omega$-invariant. For $p=0,8$ there is no IIB RR
charge and the state is automatically even under $\Omega$. In all of the cases
the net effect of the $\Omega$-projection is to eliminate the tachyon leading
to a stable brane configuration \cite{senbps}.

\subsubsection*{\it I--I\,$^\prime$ Duality}

We shall now describe the sequence of isomorphisms \eqn{botthomKO} in
the language of brane configurations, as we did in the type-II case. For
definiteness we shall take $l=2k$ to be even. The isomorphism in the first line
of \eqn{botthomKO} arises in a similar way as in the type-II case by
representing a $p-1$-brane in codimension $2k+1$ in terms of $2N$ unstable
9-branes. The induced charge takes values in the higher KO-group
\beq
\widetilde{KO}^{-1}(S^{l+1})=\pi_l\Bigl(O(2N)/[O(N)\times O(N)]\Bigr)
\label{KO1groups}\eeq
and the KO-class is determined by the rank $2N$ spinor bundle ${\cal S}_{\frac
N2}\oplus{\cal S}_{\frac N2}$ with tachyon field
\beq
T_N^{({\rm I}\,')}=T_N^{\rm(I)}\oplus T_N^{\rm(I)}+T^{(1)}\,\sigma_3\otimes I_N
\label{T2NIprime}\eeq
As in \eqn{TNI}, eq. \eqn{T2NIprime} shows that the given set of spacetime
filling 9-branes interact via tachyon fields only among themselves and not with
their mirror images. There is, however, a subtlety with this transformation
which is related to the doubling \eqn{T2NIprime} in rank of the spinor modules
due to the mirror IIB 9-branes (representing again a K-theoretic
stabilization). Under a $T$-duality transformation on $S^1$ the operator
$\Omega$ is mapped to $\Omega\cdot{\cal I}_1$, where ${\cal I}_1$ is the
geometrical operator which reflects the compactified target space direction.
The $T$-dual of the type-I theory is type-I$\,^\prime$ superstring theory which
is obtained as the $\zed_2$
orientifold projection of the type-IIA theory by $\Omega\cdot{\cal I}_1$. The
relevant K-theory is given by the Real K-group $KR(X)$
\cite{witten,atiyahreal}, i.e. the group of virtual bundles $[(E,F)]$ with an
antilinear involution on both $E$ and $F$ which commutes with the induced
action of ${\cal I}_1$ by pullback. As the type-I$\,^\prime$ theory contains
$2N$ unstable 9-branes, its D-brane charges live in the Real K-group
$\widetilde{KR}^{-1}(X)$ \cite{horava}. The relation to the present homotopy
analysis comes from the fact that if the KR-involution is taken to act
trivially on $X$, then there is a natural isomorphism
\beq
KR^{-n}(X)=KO^{-n}(X)
\label{KRKOiso}\eeq
The precise equivariant homotopy properties inherent in Real K-theory will be
discussed in more detail in the next section.

We see that the periodicity theorem \eqn{botthomKO} will in fact go beyond
describing just toroidal compactifications of the type-I theories. A
$T$-duality transformation of type-I superstring theory on $T^m$ gives a
type-II orientifold on $T^m/\zed_2$ by the operator $\Omega\cdot{\cal I}_m$
(${\cal I}_m$, along with appropriate factors of $\klein$, reflects all of the
coordinates of the compactification torus $T^m$). In the following we shall
also describe the structure of these orientifold theories using K-theoretic
properties. A general orientifold has $2^m$ orientifold planes which are the
fixed points of the $\Omega\cdot{\cal I}_m$ involution and which fill all of
the non-compact spacetime. They have dimension $9-m$ and carry $-2^{4-m}$ units
of $9-m$-brane RR-charge. We will denote them by O$(9-m)^-$. Far away from the
orientifold planes, the spacetime looks like the original $X$ along with its
mirror image under the action of the KR-involution which interchanges the two
copies. An elementary result of K-theory shows that for a trivial $\zed_2$
action on the space $X$ \cite{atiyahreal}
\beq
KR^{-n}(X\amalg X)=KR^{-n}(X\times S^{0,1})=K^{-n}(X)
\label{KRKiso}\eeq
where $S^{p,q}$ denotes the $p+q-1$ dimensional unit sphere in $\real^{p,q}$.
The theory far away from the singularities is therefore equivalent to the
original type-II theory. On the other hand, branes on top of an orientifold
plane have their generic unitary gauge symmetry broken to the real subgroup, in
accordance with the first part of the sequence \eqn{botthomKO}. As we did in
the type-II cases, we shall follow the orbit of the original stable
9-$\overline{9}$ brane pairs in the homotopy sequence, and in addition keep
track of the structure of the orientifold planes since their RR-charges will be
measured by the relevant K-theory groups. At the level of
(\ref{KO1groups},\ref{T2NIprime}), there are $N$ stable 8-$\overline{8}$ brane
pairs and two O$8^-$ planes which each carry $-8$ units of RR-charge. Tadpole
anomaly cancellation requires that the 8-branes be arranged so as to cancel out
the orientifold charge in the supersymmetric vacuum configuration. Note that
the K-group mapping between \eqn{KOSgroups} and \eqn{KO1groups}, i.e. the basic
I-I$\,^\prime$ $T$-duality, involves the homotopy groups of real symmetric
spaces.

\subsubsection*{\it m = 2}

Next we come to the first isomorphism in the second line of \eqn{botthomKO}.
This corresponds to the toroidal compactification of the type-I theory which is
$T$-dual to type-IIB superstring theory on the $T^2/\zed_2$ orientifold. There
are now four O$7^-$ planes, each with $-4$ units of 7-brane RR-charge, and $N$
7-$\overline{7}$ brane pairs which are described in terms of $2N$
9-$\overline{9}$ brane pairs, as in \eqn{tachyonN+1}. The appearence of the
unitary group in \eqn{botthomKO} is due to the following facts. The chiral
spinor modules $\Delta_2^\pm$ are complex, so that the desired map must be
taken with respect to the real spinor module $\Delta_2^+\oplus\Delta_2^-$, as
in \eqn{tachyonN+1} (see also \eqn{Bottspinmap}). This means that the relevant
homotopy is defined with respect to a unitary symmetric space. Physically, the
appearence of a unitary gauge symmetry can be understood from the analysis of
\cite{gimonpol} (see also \cite{quivers}) where the requirement of closure of
the worldsheet operator product expansion was shown to put stringent
restrictions on the actions of discrete gauge symmetries on Chan-Paton bundles.
In particular, the square of the worldsheet parity operator $\Omega$ acts on
Chan-Paton indices as
\beq
\Omega^2\,:\,~~~~|{\rm D}p;ab\rangle~\mapsto~(\Lambda_\Omega^2)^{a'}_a\,|{\rm
D}p;a'b'\rangle\,(\Lambda_\Omega^{-2})^{b'}_b=(\pm i)^{(9-p)/2}\,|{\rm
D}p;ab\rangle
\label{Om2action}\eeq
where $a,b$ are the open string endpoint Chan-Paton labels of a D$p$-brane
state of the IIB theory, and $\Lambda$ denotes the (adjoint) representation of
the orientifold group in the Chan-Paton gauge group. While the 9-branes have
the standard orthogonal subgroup projection (as required by tadpole anomaly
cancellation), \eqn{Om2action} leads to an inconsistency on 7-branes which are
therefore quantized using the unprojected unitary gauge bundles. Thus, while
the naive gauge group on the spacetime filling 9-branes is $O(2N)\times
O(2N)\subset O(4N)$, the inconsistent $\Omega$-projection on IIB 7-branes
enhances the orientifold symmetry to $U(2N)$ because from the point of view of
worldvolume gauge fields the 9-branes are indistinguishable from
$\overline{9}$-branes due to the reality property of the relevant spinor
bundles. The
requisite tachyon field $T_{2N}^{\rm(I)}$ of the corresponding virtual bundle
$[(E,F)]$ can be regarded as a section of $E\otimes\overline{F}$ and so is
required to be $\zed_2$-equivariant with respect to the orientifold projection
(this ensures that the resulting lower dimensional brane configurations are
invariant under the $\zed_2$-action), i.e. it transforms under the orientifold
group as
\beq
T_{2N}^{\rm(I)}\to\Lambda_\Omega\,T_{2N}^{\rm(I)}\,\Lambda_\Omega^{-1}
\label{tachyonOmtransf}\eeq
As shown in \cite{witten}, the tachyon vertex operator for a $p$-$\overline{p}$
brane pair acquires the phase $(\pm i)^{7-p}$ under the action of $\Omega^2$.
For the 7-branes this operator is even under $\Omega^2$, and so the eigenvalues
of the vacuum expectation value $T_{2N}^{\rm(I)0}$ are real. Thus the tachyon
field breaks the $U(2N)$ gauge symmetry down to its orthogonal subgroup
$O(2N)$, and the induced brane charge is given by the winding numbers around
the vacuum manifold $U(2N)/O(2N)$ of the IIB orientifold on $T^2/\zed_2$.

What is particularly interesting about this orientifold is that it is
equivalent to a more general, dual description in terms of $F$-theory
\cite{Ftheory}. In this description, $F$-theory on the third complex Kummer
surface $K3$ is the compactification of type-IIB superstring theory on the base
space of the elliptic fibration $K3\to\complex P^1$ with complex structure
modulus of the fiber given by the axion-dilaton modular parameter of the IIB
theory. At the special point of the $K3$ moduli space where it can be
identified with the orbifold $T^4/\zed_2$ (i.e. the degenerate $K3$ with eight
$D_2$ singularities), the fiber acquires singularities and
the base space is the orbifold limit $T^2/\zed_2$ of $\complex P^1$. The
orientifold planes split into 7-branes in this description and have some
dynamics (although they cannot support perturbative gauge fields). The
equivalence between $F$-theory on $K3$ and type-I superstring theory on $T^2$
is then a starting point for a K-theory description of $F$-theory. In this
particular compactification of $F$-theory, the relevant vacuum manifold for
tachyon condensation is $U(2N)/O(2N)$ and a K-theory subgroup of the full
compactification is $\widetilde{KR}^{-2}(Y\times T^{1,2})$ (or equivalently
$\widetilde{KO}(Y\times T^2)$ \cite{bgh} as we show in the next subsection).
Here we have used the definition $T^{1,m}=(S^{1,1})^m$. The relevant K-theory
incorporates the charges of the orientifold planes, as we will discuss further
in the following.

\subsubsection*{\it m = 3}

The second isomorphism in the second line of \eqn{botthomKO} describes the
mapping onto the I$\,^\prime$ theory on $T^3$ which is $T$-dual to the
$T^3/\zed_2$ orientifold of the type-IIA string. There are eight O$6^-$ planes
each with $-2$ units of 6-brane RR-charge. The appearence of a symplectic gauge
group follows from the mathematical fact that the complex spinor module
$\Delta_3$ is the restriction of a quaternionic Clifford module, so that the
appropriate augmentation of the spin bundles on the 9-branes is taken with
respect to the rank 4 real representation $\Delta_3\oplus\Delta_3$. This means
that there are now $4N$ unstable 9-branes which have an $Sp(2N)$ worldvolume
gauge symmetry and whose KO-theory class is represented by the spin bundle
${\cal
S}_{N}\oplus{\cal S}_{N}$ with tachyon field
\beq
T_{2N}^{({\rm I}\,')}=T_{2N}^{\rm(I)}\oplus
T_{2N}^{\rm(I)}+T^{(1)}\,\sigma_3\otimes I_{4N}
\label{tachyonm3}\eeq
This enhanced $Sp(2N)$ gauge symmetry comes from the intermediate
representation of a given type-I$\,^\prime$ $p-3$-brane in terms of
6-$\overline{6}$ brane pairs and will be more easily understood below at the
next isomorphism in terms of 5-branes. Again by $\zed_2$ equivariance the
tachyon field breaks this symmetry to its complex subgroup $U(2N)$, so that the
vacuum manifold is $Sp(2N)/U(2N)$. As shown in \cite{horava}, the codimension 3
tachyonic soliton in \eqn{tachyonm3} coincides with the usual 't Hooft-Polyakov
magnetic monopole.

Since $M$-theory compactified on a circle is the type-IIA string, a careful
analysis \cite{Mtheory} reveals that this orientifold model is equivalent to
$M$-theory defined on the orbifold limit $T^4/\zed_2$ of the 4-manifold $K3$.
This lends another clue to the puzzle of how to describe $M$-theory in
K-theoretic terms \cite{witten,horava}. The equivalence with the IIA theory
when compactified on a
circle shows that its description must then reduce to a vacuum manifold
$U(2N)/[U(N)\times U(N)]$ describing the stable soliton configurations which
represent the branes with K-group $K^{-1}(X)$, while the equivalence of the
type-IIA orientifold on $S^1/\zed_2$ to the $M$-theory orientifold on
$S^1/\zed_2$ is described in terms of the vacuum manifold $O(2N)/[O(N)\times
O(N)]$ and the KR-group $\widetilde{KR}^{-1}(Y'\times S^1)$. Now we see that
the duality between $M$-theory compactified on a $K3$ surface and
type-I$\,^\prime$ superstring theory on $T^3$ implies that this 11-dimensional
compactification has vacuum manifold $Sp(2N)/U(2N)$ and K-theory group
$\widetilde{KR}^{-3}(Y\times T^{1,3})$ (or equivalently $\widetilde{KO}(Y\times
T^3)$ \cite{bgh} as we discuss in the next subsection).

\subsubsection*{\it m = 4}

Now we come to the isomorphisms in the third line of \eqn{botthomKO}. The
superstring theory is type-I on $T^4$ which is $T$-dual to the orientifold of
type-IIB on $T^4/\zed_2$ which has 16 O$5^-$ planes each with a negative unit
of 5-brane RR charge. The rank 2 spin modules $\Delta_4^\pm$ are again the
restrictions of quaternionic Clifford modules and yield the real chiral spinor
representations $\Delta_4^\pm\oplus\Delta_4^\pm$. There are now $4N$
9-$\overline{9}$ brane pairs which determine the KO-theory class $[({\cal
S}_{2N}^+\oplus{\cal S}_{2N}^+,{\cal S}_{2N}^-\oplus{\cal S}_{2N}^-)]$ with
tachyon field
\beq
T_{4N}^{\rm(I)}=T_{2N}^{\rm(I)}\otimes(\sigma_3\oplus\sigma_3)+
T_2^{\rm(I)}\otimes I_{4N}
\label{TI4N}\eeq
where the codimension 2 tachyon field $T_2^{\rm(I)}$ is defined as in
\eqn{tachyonN+1}. The quaternionic gauge symmetry is naturally explained by
\eqn{Om2action} which shows that $\Omega^2=-1$ when acting on the $N$
5-brane-antibrane states (and also on the corresponding tachyon vertex
operator). The 5-branes must therefore be quantized using pseudo-real gauge
bundles, i.e. Chan-Paton bundles with structure group $Sp(2N)$ on the 9-branes
and on the $\overline{9}$-branes. This fact explains, via $T$-duality
transformations, the appearence of symplectic gauge groups in the $m=3$ case
above and in the cases to follow. An alternative explanation \cite{witten}
utilizes the fact that a type-I 5-brane is equivalent to an instanton on the
spacetime filling 9-branes \cite{witteninst}. The tachyon field breaks the
$SO(4N)\times SO(4N)$ gauge symmetry of the 9-$\overline{9}$ brane
configuration to the diagonal subgroup $SO(4N)_{\rm diag}$, which is then
further broken down to $Sp(2N)$ by the instanton field.

K-theory defined with pseudo-real bundles is denoted $KSp(X)$. The induced
$p-4$-brane charge of type-I superstring theory on $T^4$ is labelled by the
homotopy groups of the vacuum manifold $Sp(2N)\times Sp(2N)/Sp(2N)_{\rm diag}$
which yield the KSp-group
\beq
\widetilde{KSp}(S^{l+4})=\pi_{l+3}\Bigl(Sp(2N)\Bigr)
\label{KSpSl4}\eeq
In the language of the orientifold theory, the replacement of the KR-involution
by a $\zed_4$ generator defines the pseudo-Real K-group $KH(X)$ \cite{kh}
appropriate to orientifold theories with symplectic Chan-Paton
bundles.\footnote{Strictly speaking, the $\Omega\cdot{\cal I}_m$ orientifold
projection should be accompanied by the action of the operator
$(-1)^{\frac12(9-p)(8-p)F_{\rm L}}$ to preserve the $\zeds_2$-equivariant
structure in the homotopy sequence. The usage of the KH-theory generalization
of the orientifold models has been discussed in \cite{gukov,hori}.} The present
model is dual to a conventional orbifold of the type-IIA theory on $T^4/\zed_2$
and its moduli space coincides with that of IIA strings on $K3$ \cite{bsv}.

\subsubsection*{\it m = 5}

The situation for type-I$\,^\prime$ theory on $T^5$ introduces a new chain of
dualities which may be attributed to the fact that the orientifold planes now
begin acquiring fractional RR-charges. This theory is $T$-dual to the
$T^5/\zed_2$ orientifold of type-IIA superstring theory with 32 O$4^-$ planes
each of
fractional magnetic charge $-\frac12$. This new duality chain can also be
understood in the K-theory language from the change of nature of the Chan-Paton
spinor bundles on the 9-$\overline{9}$ branes at the $m=4$ case above. Again
the spinor module $\Delta_5$ is quaternionic in origin and yields the eight
dimensional real spinor representation $\Delta_5\oplus\Delta_5$. This is
consistent with the physical expectations from $T$-duality of the pseudo-real
vacuum manifold associated with $8N$ unstable 9-branes carrying the Chan-Paton
bundle ${\cal
S}_{2N}\oplus{\cal S}_{2N}$ (with tachyon field $T_{4N}^{({\rm I}\,')}$ defined
analogously to \eqn{tachyonIIA}) which appears in the second isomorphism of the
third line of \eqn{botthomKO}. A candidate dual theory is the $K3\times S^1$
compactification of type-IIB superstring theory, which may then be conjectured
to be dual to a $T^5/\zed_2$ orientifold of $M$-theory \cite{IIBK3}. This
chiral $M$-theory orientifold has no D-branes but rather M5-branes on which
M2-branes can end. The chiral lift of the $m=5$ string orientifold should then
be related to various equivalent K-groups given by $\widetilde{KR}^{-5}(Y\times
T^{1,5})$, $\widetilde{KO}(Y\times T^5)$, $\widetilde{KSp}(Y'\times S^1)$ and
$\widetilde{KH}^{-1}(Y'\times S^{1,1})$. Along with the recent construction
\cite{yi} of M2-branes as tachyonic-type solitons in the worldvolume of
M5-$\overline{\rm M5}$ brane configurations, this could shed more light on the
interpretation of $M$-theory using K-theory (some related results can also be
found in \cite{matrix}).

\subsubsection*{\it m = 6}

The type-I superstring on $T^6$ is $T$-dual to the IIB string on $T^6/\zed_2$
which has 64 O$3^-$ planes each of fractional RR-charge $-\frac14$ filling the
non-compact space. The $N$ 3-$\overline{3}$ brane pairs, which are described as
the tachyonic kinks of $8N$ 9-$\overline{9}$ brane pairs, have the same fate
according to \eqn{Om2action} as the 7-branes, and hence the 9-brane gauge
symmetry is the unprojected $U(8N)$. Algebraically this owes to the fact
that the spinor modules $\Delta_6^\pm$ are complex and thus form the real
spinor representation $\Delta_6^+\oplus\Delta_6^-$. The corresponding KO-group
element is thus $[({\cal S}_{4N}^+\oplus\overline{\cal S}_{4N}^-,\overline{\cal
S}_{4N}^+\oplus{\cal S}_{4N}^-)]$, and $\zed_2$ equivariance of the tachyon
field implies that it breaks the $U(8N)$ gauge symmetry to its pseudo-real
subgroup $Sp(4N)$. Again this model can be described as the limit of an
$F$-theory compactification on the orbifold $T^8/\zed_2$ which has terminal
singularities and is not the limit of any suitable smooth 8-manifold. This
$F$-theory compactification should thus reduce to the vacuum manifold
$U(8N)/Sp(4N)$ around which the tachyonic windings give rise to K-theory
classes in $\widetilde{KR}^{-6}(Y\times T^{1,6})$ (or the various other
equivalent K-groups).

\subsubsection*{\it m = 7}

The $T^7/\zed_2$ orientifold of the type-IIA theory has 128 O$2^-$ planes each
of electric charge $-\frac18$. The spinor module $\Delta_7$ is real, so that
the $16N$ unstable 9-branes support the spinor bundle ${\cal S}_{4N}\oplus{\cal
S}_{4N}$ and have a real $O(16N)$ gauge symmetry. This reality condition arises
from the intermediate 2-$\overline{2}$ brane pairs and again will be explained
at the next isomorphism in terms of D-strings. The tachyon field breaks the
$O(16N)$ gauge symmetry to the unitary subgroup $U(8N)$, as follows from
equivariance with respect to the orientifold group. This model can be mapped to
the compactification of $M$-theory on $T^8/\zed_2$, from which we can identify
another limiting vacuum manifold $O(16N)/U(8N)$ and K-group
$\widetilde{KR}^{-7}(Y\times T^{1,7})$.

\subsubsection*{\it m = 8}

The final isomorphism in the sequence \eqn{botthomKO} maps us into the type-I
theory on $T^8$ which is $T$-dual to the type-IIB orientifold on $T^8/\zed_2$
that has 256 O$1^-$ planes each of RR-charge $-\frac1{16}$. The vacuum manifold
is now the real orthogonal group $O(16N)$ because the intermediate IIB
D-strings have the usual orthogonal projection according to \eqn{Om2action}.
This is consistent with the fact that the rank 8 chiral spinor modules
$\Delta_8^\pm$ are real, and thus the $16N$ 9-$\overline{9}$ brane pairs
support Chan-Paton bundles which each transform under the
$\Delta_{2k+8}^+\oplus\Delta_{2k+8}^-$ spinor representation of the transverse
rotation group, similarly to the case we started with. This
orientifold is equivalent to the ordinary orbifold compactification of type-IIA
superstring theory on $T^8/\zed_2$, which has fundamental strings condensed in
the supersymmetric vacuum. This latter property and its K-theory origin may
again be combined with the analysis of \cite{yi} which shows that the
solitonic description of an M2-brane from the annihilation of a coincident pair
of M5-$\overline{\rm M5}$ branes reduces (upon compactification on $S^1$) to a
fundamental string stretched between an annihilating pair of D4-$\overline{\rm
D4}$ branes in the IIA theory. The closure of the Bott periodicity sequence
\eqn{botthomKO} at this stage implies that there is no new physics at lower
compactifications, as is indeed precisely the case.

\subsubsection*{\it Hopf Maps}

The strong Bott periodicity isomorphisms for real and pseudo-real K-theory take
the form
\bea
\widetilde{KO}^{-n-8}(X)=\widetilde{KO}^{-n}(X)~~~~~~&,
&~~~~~~\widetilde{KR}^{-n-8}(X)=\widetilde{KR}^{-n}(X)
\label{bottRealKn}\\\widetilde{KSp}^{-n-4}(X)=\widetilde{KO}^{-n}(X)
{}~~~~~~&,&~~~~~~\widetilde{KH}^{-n-4}(X)=\widetilde{KR}^{-n}(X)
\label{bottrealKn}\eea
They can again be deduced from the periodicity relations
$C_{l+8}=C_l\otimes\real(16)$ and $C_{l+4}=C_l\otimes_\reals\quater(2)$ of the
corresponding Clifford algebras \cite{spingeom}. As mappings on K-theory
classes they can be represented via cup products involving spinor bundles, as
in \eqn{Bottspinmap}, or equivalently by using Hopf fibrations as
in (\ref{alphaiso},\ref{altBottmap}). The isomorphism of KO-groups in
\eqn{bottRealKn} comes from taking the cup product of an element of
$\widetilde{KO}^{-n}(X)$ with the generator $[{\cal N}_\reals]-[I^7]$ of
$\widetilde{KO}(S^8)=\zed$, where ${\cal N}_\reals$ is the rank 7 Hopf bundle
over $\real P^8$ associated with the real Hopf fibration $S^{15}\to S^8$. The
corresponding mapping on the Real K-theory groups is obtained via the natural
Real structure on the complex Hopf bundle over $\complex P^7$
\cite{atiyahreal}. This shows that the construction of a $p$-brane in terms of
$p+8$-branes (e.g. a type-I D-particle from 8-$\overline{8}$ brane pairs) is
determined by a D-string solitonic configuration which gives another explicit
physical realization of the $spin(8)$ instanton.\footnote{Note that for the
orientifold models, the equivariance condition \eqn{tachyonOmtransf} on the
tachyon field and a similar one on the worldvolume gauge fields implies that
the topological defects arising from the Hopf
fibrations are always equivariant versions of these solitons.} The
corresponding eight dimensional non-trivial gauge connections, and the
associated spinor structures, may be found in \cite{hopf8}. This minimizing
solution of the eight dimensional Euclidean Yang-Mills equations satisfies a
generalized duality condition with respect to the topological 4-form $F\wedge
F$ constructed from the associated field strength. This identifies the explicit
form of the worldvolume gauge fields living on the $p+8$-$\overline{p+8}$ brane
pair required to produce the finite energy solitonic $p$-brane configuration as
\cite{hopf8}
\beq
A_i(x)=-2i\,\sum_{j=1}^8\Gamma_{ij}\,\frac{x^j}{(1+|x|^2)^2}
\label{8gaugefield}\eeq
where $\Gamma_{ij}$ are the generators of $spin(8)$. These gauge field
configurations are $spin(9)$ symmetric (thereby preserving the manifest
spacetime symmetries) and carry unit topological charge.

Similarly, the isomorphism of pseudo-real K-groups in \eqn{bottrealKn} comes
from taking the cup product with the class of the rank 2 instanton bundle
${\cal N}_\quaters$ associated with the pseudo-real Hopf fibration $S^7\to S^4$
\cite{trautman}, i.e. the holomorphic vector bundle of rank 2 over $\complex
P^3$. Thus the relationship between a BPS $p$-brane and a BPS $p+4$-brane is a
5-brane soliton which may be identified with an $SU(2)$ Yang-Mills instanton
field. This descent relation was noted in \cite{seninst} in the case of a
type-I D-string in the worldvolume of a 5-$\overline{5}$ brane pair. The
worldvolume gauge symmetry is $SU(2)\times SU(2)$ and the tachyon field
transforms in its ${\bf2}\otimes{\bf\overline{2}}$ representation. The
$\Omega$-projection identifies the vacuum manifold of the 5-brane configuration
as $SU(2)=Sp(1)$, and the finite energy static string solution in the
corresponding 5+1 dimensional field theory has asymptotic boundary $S^3$. By
choosing the asymptotic form of the tachyon field as in \eqn{puregauge}, the
topological stability of the string is guaranteed by the homotopy group
$\pi_3(SU(2))=\zed$. Arranging the asymptotic form of the gauge field on the
5-brane as in \eqn{puregauge} (and that on the $\overline{5}$-brane to vanish),
the string soliton carries 1 unit of instanton number which is a source of
D-string charge in type-I string theory \cite{witteninst}. These arguments
agree precisely with the general homotopy analysis of the $m=4$ case above.

We see that the descent relations in type-II and type-I superstring theory give
natural realizations of all three higher Hopf fibrations. The elementary Hopf
fibration $S^1\to S^1$ with discrete fiber $\zed_2$ arises in the construction
of a type-I non-BPS brane as a codimension 1 kink of a brane-antibrane pair,
e.g. the type-I D-particle from a D-string anti-D-string pair
\cite{witten,seninst}. The double cover of $S^1$ corresponds to the pair of
D-strings, and the winding number of the tachyon field is labelled by the
homotopy group $\pi_0(\zed_2)=\zed_2$ of the fiber corresponding to the
discrete gauge transformation $T^{(1)}\to-T^{(1)}$. This agrees with the
degenerate situation of describing an 8-brane in terms of a single
9-$\overline{9}$ brane pair \cite{horava}, whereby the vacuum manifold is
$O(1)=\{\pm\,T^{(1)0}\}$ and one of the 9-branes carries a $\zed_2$ Wilson
line. The cup product with the generator of $\widetilde{KO}(S^1)=\zed_2$ then
achieves the mapping of $\zed_2$-valued KO-theory classes. Therefore, the four
fundamental Hopf fibrations are responsible for the complete spectrum of
D-brane charges in type-II and type-I superstring theory.

These cup products also shed light on the appearence of the rich spectrum of
brane charges in the type-I theories as compared to the type-II theories. Since
any real vector bundle may be regarded as complex, and any complex one as
quaternionic, there is a natural homomorphism between the K-groups of the
type-I and type-II theories,
\beq
\widetilde{KO}^{-n}(X)~\longrightarrow~\widetilde{K}^{-n}(X)~
\longrightarrow~\widetilde{KSp}^{-n}(X)
\label{KSPOmap}\eeq
As an example of this map, consider the generator of $\widetilde{K}(S^4)=\zed$,
which is the pseudo-real instanton bundle described above. To realize it as a
generator of $\widetilde{KO}(S^4)=\zed$, which labels type-I 5-brane charge, it
must be embedded in the orthogonal structure group as $SO(4)=SU(2)\times SU(2)$
which then doubles the 5-brane charge \cite{witten}. Thus the natural map
between the integer-valued K-groups $\widetilde{K}(S^4)$ and
$\widetilde{KO}(S^4)$ is multiplication by 2, $\zed\to2\zed$. Upon forming
cosets in the ring
structure of the KO-groups (see \eqn{absisoKO}), we see how extra $\zed_2$
charged objects arise in the spectrum of the type-I theories (see
\cite{spingeom} for the mathematical details).

\newsubsection{Duality Transformations}

We shall now describe the various transformations among the dual theories
described in the previous section. We will first discuss the duality between
the type-I and type-I$\,^\prime$ theories. As in subsection 2.2 we may readily
compute
\bea
\widetilde{KO}(Y\times
S^1)&=&\left(\widetilde{KO}^{-1}(Y)\oplus\zed_2\right)
\oplus\widetilde{KO}(Y)\label{KOYS1}\\\widetilde{KR}^{-1}
(Y\times S^{1,1})&=&\left(\widetilde{KO}(Y)\oplus\zed\right)
\oplus\widetilde{KO}^{-1}(Y)
\label{KRYS1}\eea
where we have used \eqn{KRKOiso} and the fact that the
orientifold group does not act on the non-compact space $Y$. Modulo the usual
discrete groups of gauge transformations, the K-groups \eqn{KOYS1} and
\eqn{KRYS1} of the type-I and type-I$\,^\prime$ theories are the same. The
cyclic subgroup in \eqn{KOYS1} comes from \eqn{absisoKO}, while the integer
subgroup in \eqn{KRYS1} comes from the identity \cite{atiyahreal}
\beq
\widetilde{KR}^{-n}(S^{p,q})=
\widetilde{KR}^{q+1-n-p}(S^{1,0})=\widetilde{KO}^{q+1-n-p}(S^0)
\label{KRSpq}\eeq
It arises in exactly the
same way as in the type-II case, i.e. from the tachyon field \eqn{TNI} which is
the large gauge transformation of the $O(N)$ gauge bundle over the sphere
$S^l$. The cyclic subgroup of \eqn{KOYS1} comes from the $\zed_2$ gauge
transformation in the degenerate codimension 1 vacuum manifold, as explained
above. Thus the duality transformations between the type-I and I$\,^\prime$
theories correctly account for the relevant changes of gauge field
configurations living on the brane worldvolumes, just as in the type-II
theories.\footnote{For an alternative construction of the $\zeds_2$ Wilson
lines, see \cite{bgh}.}

Again at the level of {\it unreduced} K-theory, the two K-groups coincide. By
iterating \eqn{KOYS1} and \eqn{KRYS1}, we find that the same is true of the
higher dimensional toroidal compactifications demonstrating the explicit mod
$\zed$ equivalence of the type-I and type-I$\,^\prime$ models, or equivalently
of their $T$-dual type-II orientifold theories,
\bea
KO(Y\times T^m)&=&\bigoplus_{n=0}^mKO^{-n}(Y)^{\oplus{m\choose
n}}\label{KOTm}\\KR^{-m}(Y\times
T^{1,m})&=&\bigoplus_{n=0}^mKO^{n-m}(Y)^{\oplus{m\choose n}}
\label{KRTm}\eea
The decompositions of K-groups in (\ref{KOTm},\ref{KRTm}) contain the relevant
degeneracies of wrapped branes around the cycles of $T^m$, which in the case of
\eqn{KRTm} identifies the distribution of brane charges over the $2^m$
orientifold planes. As in subsection 2.2, each subgroup
$KO^{n-m}(Y)^{\oplus{m\choose n}}$ is generated by a descendent tachyon field
as one cycles through the periodicity maps in \eqn{botthomKO}. Upon writing the
isomorphism between \eqn{KOTm} and \eqn{KRTm} in terms of reduced K-groups, we
obtain winding numbers of the tachyon fields corresponding to large gauge
transformations around the various cycles. In the type-II cases, these winding
numbers indicated the precise degeneracy $2^{m-1}$ associated with the fact
that the branes transformed under the spinor representation of the target space
duality group. In the present
case, we do not know the full duality group of a generic type-II
orientifold, much less the representation that the brane charges carry. The
decompositions into reduced groups in \eqn{KRTm} should be a clue as to what
the appropriate group theoretic properties are of the target space dualities in
this case. The results are summarized in table 1.

\begin{table}
\begin{center}
\begin{tabular}{|c|c|c|l|} \hline
\ $m$ \ & \ Dual theory\ & \ Vacuum manifold\ & $\widetilde{KR}^{-m}(Y\times
T^{1,m})$\\ \hline\hline
0 & Type-I & $O(N)$ & $\widetilde{KO}(X)$\\ \hline
1 & Type-I$\,^\prime$ & $\frac{O(2N)}{O(N)\times O(N)}$ &
$\widetilde{KO}(Y)\oplus\widetilde{KO}^{-1}(Y)\oplus\zed$\\ \hline & &  &
$\widetilde{KO}(Y)\oplus\widetilde{ KO}^{-1}(Y)^{\oplus
2}$\\\raisebox{1.5ex}[0pt]{2} & \raisebox{1.5ex}[0pt]{$F$-theory on $K3$} &
\raisebox{1.5ex}[0pt]{$\frac{U(2N)}{O(2N)}$} &
$\oplus\,\widetilde{KO}^{-2}(Y)\oplus\zed\oplus(\zed_2)^{\oplus 2}$\\ \hline{}
& {} & {} & $\widetilde{KO}(Y)\oplus
\widetilde{KO}^{-1}(Y)^{\oplus 3}$\\
3 & $M$-theory on $K3$ & $\frac{Sp(2N)}{U(2N)}$ & $\oplus\,
\widetilde{KO}^{-2}(Y)^{\oplus 3}\oplus \widetilde{KO}^{-3}(Y)$\\
{} & {} & {} & $\oplus\,\zed\oplus(\zed_2)^{\oplus 6}$\\ \hline
{} & {} & {} & $\widetilde{KO}(Y)\oplus \widetilde{KO}^{-1}(Y)^{\oplus 4}$\\
4 & IIA on $K3$ &  $Sp(2N)$ & $\oplus\,\widetilde{KO}^{-2}(Y)^{\oplus 6}\oplus
\widetilde{KO}^{-3}(Y)^{\oplus 4}$\\
{} & {} &  {} & $\oplus\,\widetilde{KO}^{-4}(Y)\oplus\zed\oplus
(\zed_2)^{\oplus 10}$\\ \hline{} & {} & {} &
$\widetilde{KO}(Y)\oplus \widetilde{KO}^{-1}(Y)^{\oplus 5}$\\
 &  &  &
$\oplus\,\widetilde{KO}^{-2}(Y)^{\oplus 10}\oplus
\widetilde{KO}^{-3}(Y)^{\oplus 10}$\\
\raisebox{1.5ex}[0pt]{5} & \raisebox{1.5ex}[0pt]{IIB on $K3\times S^1$} &
\raisebox{1.5ex}[0pt]{$\frac{Sp(4N)}{Sp(2N)\times Sp(2N)}$} &
$\oplus\,\widetilde{KO}^{-4}(Y)^{\oplus 5}\oplus \widetilde{KO}^{-5}(Y)$\\
{} & {} & {} & $\oplus\,\zed^{\oplus 6}\oplus(\zed_2)^{\oplus 15}$\\ \hline
{} & {} & {} & $\widetilde{KO}(Y)\oplus
\widetilde{KO}^{-1}(Y)^{\oplus 6}$\\
 &  &  & $\oplus\,\widetilde{KO}^{-2}(Y)^{\oplus 15}\oplus
\widetilde{KO}^{-3}(Y)^{\oplus 20}$\\
\raisebox{1.5ex}[0pt]{6} & \raisebox{1.5ex}[0pt]{$F$-theory on $T^8/\zed_2$} &
\raisebox{1.5ex}[0pt]{$\frac{U(8N)}{Sp(4N)}$} &
$\oplus\,\widetilde{KO}^{-4}(Y)^{\oplus 15}\oplus
\widetilde{KO}^{-5}(Y)^{\oplus 6}$\\
{} & {} & {} & $\oplus\,\widetilde{KO}^{-6}(Y)\oplus\zed^{\oplus 16}\oplus
(\zed_2)^{\oplus 21}$\\ \hline{} & {} & {}& $\widetilde{KO}(Y)\oplus
\widetilde{KO}^{-1}(Y)^{\oplus 7}$\\ &  &  & $\oplus\,
\widetilde{KO}^{-2}(Y)^{\oplus 21}\oplus \widetilde{KO}^{-3}(Y)^{\oplus35}$\\
7 & $M$-theory on  $T^8/\zed_2$ & $\frac{O(16N)}{U(8N)}$ &
$\oplus\,\widetilde{KO}^{-4}(Y)^{\oplus 35}\oplus
\widetilde{KO}^{-5}(Y)^{\oplus 21}$\\
{} & {} & {} & $\oplus\,\widetilde{KO}^{-6}(Y)^{\oplus 7}\oplus
\widetilde{KO}^{-7}(Y)$\\{} & {} & {} & $\oplus\,\zed^{\oplus
36}\oplus(\zed_2)^{\oplus 28}$\\ \hline{} & {} & {} & $\widetilde{KO}(Y)\oplus
\widetilde{KO}^{-1}(Y)^{\oplus 8}$\\
 &  &  & $\oplus\,\widetilde{KO}^{-2}(Y)^{\oplus
28}\oplus \widetilde{KO}^{-3}(Y)^{\oplus 56}$\\
8 & IIA on $T^8/\zed_2$ & $O(16N)$ & $\oplus\,\widetilde{KO}^{-4}(Y)^{\oplus
70}\oplus \widetilde{KO}^{-5}(Y)^{\oplus 56}$\\
{} & {} & {} & $\oplus\,\widetilde{KO}^{-6}(Y)^{\oplus 28}\oplus
\widetilde{KO}^{-7}(Y)^{\oplus 8}$\\{} & {} & {} &
$\oplus\,\widetilde{KO}^{-8}(Y)\oplus\zed^{\oplus 71}\oplus(\zed_2)^{\oplus
36}$\\ \hline
\end{tabular}
\end{center}
\caption{\baselineskip=12pt {\it Type-II orientifolds on spacetimes
$X=Y\times T^{1,m}$. The general dual orbifold model in each case is listed
along with the corresponding vacuum manifold for tachyon condensation in the
worldvolume of $2^{[\frac m2]+1}N$ spacetime filling 9-branes. The last column
represents the distribution of brane charges over the various orientifold
planes and their relevant multiplicities according to some representation of
the target space duality group of the orientifold theory.}}
\end{table}

\subsubsection*{\it Orbifold Dualities}

As discussed in the previous subsection, all of the type-II orientifold string
theories are conjectured to be dual to more general conventional orbifold
theories (see the second column of table 1). Although we cannot test these
relations in general, we can make a heuristic analysis for the $m=4,5,8$ cases
in table 1. The relevant K-groups for the type-IIA orbifolds on $T^m/\zed_2$
are given by the equivariant cohomology $K^{-1}_{\zeds_2}(Y\times T^{1,m})$.
These groups can be computed as in subsection 2.3 using the six term exact
sequence \eqn{sixterm}, with the results
\bea
K^{-1}_{\zeds_2}(Y\times T^{1,4})&=& \left( K^{-1}(Y)^{\oplus 16} \otimes
R[\zed_2]\right)\oplus K^{-1}(Y)\label{IIAT4orb}\\
K^{-1}_{\zeds_2}(Y\times T^{1,8})&=& \left( K^{-1}(Y)^{\oplus 256} \otimes
R[\zed_2]\right)\oplus K^{-1}(Y)
\label{IIAT8orb}\eea
For the type-IIB orbifold on $S^1\times T^4/\zed_2$, we use in addition the
decompositions (\ref{wKXciso},\ref{K1Xciso}) to get
\bea
\widetilde{K}_{\zeds_2}(Y\times S^1\times
T^{1,4})&=&\left(\widetilde{K}(Y)^{\oplus16}\oplus
K^{-1}(Y)^{\oplus16}\oplus\zed^{\oplus15}\right)\otimes R[\zed_2]\nn\\&
&\oplus\,K^{-1}(Y)\oplus\widetilde{K}(Y)
\label{IIBT4orb}\eea

Let us compare the $m=4$ line of table 1 with the complex K-group
\eqn{IIAT4orb} which represents the spectrum of D-brane charges in the orbifold
limit of the IIA compactification on $K3$. Both decompositions contain the
multiplicities of brane charges localized on the 16 fixed point 5-planes of the
given involution. The orientifold K-groups \eqn{KRTm} contain in addition the
tachyon field winding numbers around cycles of $T^4$ as well as the appropriate
windings around the various vacuum manifolds as dictated by the general
homotopy analysis of subsection 3.1. On the other hand, the orbifold K-group
\eqn{IIAT4orb} contains the mirror image brane charges on the orbifold planes
along with the contribution from the unwrapped IIA brane configurations
(represented by the second $K^{-1}(Y)$ direct summand). The natural map
\eqn{KSPOmap} between real and complex K-groups shows how the IIA orbifold
charges correspond to precisely the orientifold charges from a given tachyon
configuration, along with the appropriate multiplicity of 2 as discussed at the
end of the previous subsection. The remnant large gauge symmetry of the
orientifold charges are then represented by unwrapped orbifold charges. This
provides a new relationship between the given type-II orientifold theory and
its dual. Of course, this heuristic comparison only holds at the level of the
orbifold limit of the $K3$ moduli space of the IIA orbifold theory. A more
precise analysis of this duality should make a comparison with the full $K3$
compactification.

The group \eqn{IIAT4orb} can be seen to account for the usual BPS branes of the
IIA orbifold theory \cite{sendescent,bg}. This spectrum contains fractional
D-particles of unit charge with respect to the twisted $U(1)$ RR gauge fields
at the orbifold planes, with the correct multiplicity of 4 arising from the
possible bulk and twisted charges of the states at each plane giving a total of
64 such states. The spectrum also contains wrapped D2-branes around
non-vanishing supersymmetric cycles, and D4-branes which wrap around the entire
compact space. However, it is less clear how to identify the spectrum of stable
non-BPS configurations directly from the decomposition \eqn{IIAT4orb}. For
instance, the IIA theory on $T^4/\zed_2$ should contain a $\zed_2$-charged
non-BPS 4-brane which comes from the D-particle in the dual type-I string
theory. The spectrum of $\zed_2$-charges in general can only be identified
under the natural homomorphism between \eqn{IIAT4orb} and table 1, so
that we can take the latter groups to represent the full brane spectrum of the
orbifold theory. The problem can be traced back to the fact that the duality
map in the present case involves an intermediate $S$-duality transformation
\cite{bg}, whose description in the K-theoretic formalism is at present not
known \cite{witten,gukov}. Other non-BPS states in the spectrum of the present
model include D-particles stuck at the fixed point planes, and D-strings
stretched between pairs of orbifold fixed points with the same magnitude of
charge as
those of the fractional BPS D-particles \cite{bg}.

A similar comparison holds between the last line of table 1 and the complex
K-group \eqn{IIAT8orb} which represents the type-IIA orbifold compactification
on $T^8/\zed_2$. The duality between the $m=5$ line of table 1 and
\eqn{IIBT4orb}, which represents the orbifold limit of the IIB compactification
on $K3\times S^1$, is more involved because the latter group contains
extra unwrapped and wrapped brane charges. The 32 O$4^-$ planes of the IIB
orientifold have brane charges distributed according to table 1. On the other
hand, the brane charges localized at the 16 fixed point 5-planes of the IIB
orbifold (along with the mirror images) split evenly into two sets of type-II
charges corresponding to wrapped and unwrapped configurations around the extra
$S^1$. Now the mapping between the two K-groups according to \eqn{KSPOmap}
matches the symmetrical splitting of the KO-group decomposition for $m=5$, and
one may take the K-groups of table 1 to represent the non-BPS configurations of
these dual models. It would be interesting to test this matching by explicit
string theoretical constructions.

\newsubsection{$(-1)^{F_{\rm L}}$ Transformations}

The generalization of the action of the Klein operator $\klein$ on the type-I
theories and on the type-IIB orientifolds can be deduced from their
relationships with the type-II theories. Let us start from the type-I theory,
regarded as the quotient of type-IIB superstring theory by worldsheet parity
$\Omega$. Regarding the type-IIA theory as the quotient of IIB by
$\klein$, the type-I$\,^\prime$ theory may then be obtained as the quotient of
IIB by the involution $\Omega\cdot\klein\cdot{\cal I}_1$. The Grothendieck
group $KR_\pm(X)$ of virtual complex vector bundles with such an involution is
a generalization of the Hopkins K-groups to the category of Real vector bundles
over $X$ \cite{witten,gukov}. Thus the action on K-theory by the $\klein$
projection on the type-I theory should induce a map
\beq
\widetilde{KO}(X)~\longrightarrow~\widetilde{KR}_\pm(X)
\label{KOKRpmmap}\eeq
The problem we encounter at this stage is a mathematical one. The theory of
Hopkins groups $K_\pm(X)$ has not been investigated much in the literature,
much less its Real generalization. In particular, a product formula such as
that given by the right-hand side of \eqn{Kequivmap} is not known in this case.
In \cite{gukov} it was suggested that an analog of the Hopkins formula could be
\beq
KR_\pm(X)=KR_{\zeds_2}(X\times\real^{1,1})
\label{KRpmprod}\eeq
where the cyclic group acts as the product of the action of ${\cal I}_1$ on $X$
and an orientation reversing symmetry of the real space $\real^{1,1}$.

This relationship exemplifies the fact that duality transformations and
orbifold operations do not always commute \cite{senorb}, in this case within
the various interrelationships between the type-I and type-II theories. Instead
of the mapping \eqn{KOKRpmmap}, there is a more natural candidate which comes
about in analogy with the $T$-duality transformations of the type-I theories.
One could consider the action of $\klein$ directly on the type-I$\,^\prime$
theory, thereby obtaining the map \eqn{Kequivmap} into Real virtual bundles.
The relevant K-group for the operation of modding out the type-I$\,^\prime$
theory $m$ times by $\klein$ is then
\beq
\widetilde{KR}^{-m}_{\zeds_2}(X\times\real^{0,m})=
\left(\widetilde{KR}^{-m}(X)\otimes R[\zed_2]\right)\oplus\widetilde{KR}(X)
\label{KRkleinm}\eeq
with a trivial $\zed_2$ action on the spacetime $X$. The decomposition
\eqn{KRkleinm} follows using the methods of subsection 2.3 applied to the
KR-groups and the suspension isomorphism \cite{spingeom}
\beq
KR^{-n}(X\times\real^{p,q})=KR^{q-p-n}(X)
\label{KRsusp}\eeq
for Real K-theory. The projection onto the first direct summand in
\eqn{KRkleinm}, representing the states which survive the $\klein$ projections,
can be carried out explicitly on the invariant brane-antibrane configurations
as in subsection 2.3. The other summand $\widetilde{KR}(X)$ then always
represents the orientifold states projected out by the $\klein$ involution. The
subsequent $\klein$ projections now take us through the entire sequence of
type-II
orientifolds described in subsection 3.1. It would be
interesting to test the mappings described here directly using explicit string
theoretical constructions (for example, a boundary state calculation along the
lines of \cite{daspark}).

\newsubsection{Summary}

Again we may succinctly summarize the descent relations among type-I branes and
those of type-II orientifolds by a diagram representing the various mappings on
K-groups:
\beq\new{\begin{array}{ccllllc}
KO(X)&{\buildrel\beta\over\longrightarrow}&KO^{-1}(X)&{\buildrel\beta
\over\longrightarrow}&\!\!\!\!\!\!\!\!\cdots&\!\!\!\!\!\!\!\!
{\buildrel\beta\over\longrightarrow}&\!\!\!\!\!\!\!\!
KO^{-8}(X)\\& & &\!\!\!\!\!\!\!\!&\!\!\!\!\!\!\!\!&\!\!\!\!\!\!\!\!&\!\!
\!\!\!\!\!\!\parallel\\{\scriptstyle\klein}\downarrow&\searrow&\!\!\!\!\!\!\!
{\scriptstyle KO(Y\times S^1)}&\!\!\!\!\!\!\!\!&\!\!\!\!\!\!\!\!&
\!\!\!\!\!\!\!\!&\!\!\!\!\!\!\!\!KO(X)\\& &{\scriptstyle=KR^{-1}(Y\times
S^{1,1})}& & & &\\& & & & & &\\KR_{\zeds
_2}^{-1}(X\times\real^{0,1})&{\buildrel{\beta\circ\Pi_1}\over\longrightarrow}
&KR^{-1}(X)& & & &\\& & & & & &\\& &{\scriptstyle KO(Y\times T^2)}& & & &\\
{\scriptstyle\klein}\downarrow& &{\scriptstyle=KR^{-2}(Y\times T^{1,2})}
&\!\!\!\!\!\!\!\searrow& & &\\&
 & & & & &\\KR_{\zeds_2}^{-2}(X\times\real^{0,2})& &{\buildrel
{\beta\circ\Pi_1}\over\longrightarrow}&KR^{-2}(X)& & &\\& & & & & &\\&
& &{\scriptstyle KO(Y\times T^3)}& & &\\{\scriptstyle\klein}\downarrow& &
&{\scriptstyle=KR^{-3}(Y\times T^{1,3})}&\!\!\!\!
\!\!\!\searrow& &\\& & & & & &\\\vdots& & & &\ddots& &\\& & & &
{\scriptstyle KO(Y\times T^8)}& &\\{\scriptstyle\klein}\downarrow& &
& &{\scriptstyle=KR(Y\times T^{1,8})}&\!\!\!\!\!\!\!\searrow& \\& & & & & &
\\KR_{\zeds_2}^{-8}(X\times\real^{0,8})& &
&{\buildrel\Pi_1\over\longrightarrow}& &KR(X)&\end{array}}
\label{KOKRdiag}\eeq
Again $\beta$ is the Bott periodicity isomorphism representing tachyon
condensation in codimension 1, except that now in general it leads to a stable
brane configuration (non-trivial K-theory class). In particular, $\beta$
realizes a type-I non-BPS $p-1$-brane as a kink in the worldvolume of a
$p$-$\overline{p}$ pair. The map $\beta^4$ acts on KO-groups according to
\beq
\beta^4\,KO(X)=KSp(X)
\label{beta4KOX}\eeq
and it realizes a BPS $p-4$-brane as an $SU(2)$ Yang-Mills instanton in the
worldvolume of four $p$-$\overline{p}$ pairs. The map $\beta^8$ realizes a
$p-8$-brane as a $spin(8)$ instanton in the worldvolume of 16
$p$-$\overline{p}$ pairs. The projections $\Pi_1$ are as
before. Note that the statements about KR-groups above can be translated into
ones about KO-groups upon taking the KR-involution to act trivially on the
spacetime $X$.

\newsection{Orientifold Symmetries}

One subtlety in our description of type-II orientifolds in the previous
section is that the homotopy classification of KR-theory requires a slightly
refined definition. This is in turn related to a more general periodicity
property of Real K-theory. In this final section we will describe these general
symmetries of the orientifold models in some detail, thereby exposing the rich
internal symmetries predicted by the K-theory formalism.

\newsubsection{Periodicity in Real K-theory}

The definition of the higher KR-groups that we have used thus far has been done
with respect to suspensions whereby the KR-involution acts trivially on the
sphere. There is a more general class of Real K-groups which are defined by the
double index suspensions
\beq
KR^{p,q}(X)=KR(\Sigma^{p,q}X)~~~~~~{\rm
with}~~\Sigma^{p,q}X=X\wedge\real_+^{p,q}
\label{KRpqdef}\eeq
With this definition we have
\beq
KR^{-n}(X)=KR^{n,0}(X)
\label{KRdiff}\eeq
Taking the cup product with the class of the complex Hopf line bundle ${\cal
N}_\complexs$ over $\complex P^1$ with its natural Real structure (given by the
anti-linear complex conjugation involution) induces the strong Bott periodicity
isomorphism \cite{atiyahreal}
\beq
KR^{p+1,q+1}(X)=KR^{p,q}(X)
\label{KRbottper}\eeq
showing that in fact $KR^{p,q}(X)=KR^{q-p}(X)$. Identifying $\real^{1,1}$ as
the space $\complex$ with complex conjugation, this $(1,1)$ periodicity may be
cast in the form of a suspension isomorphism
\beq
KR(X)=KR(X\times\complex)
\label{KRsuspiso}\eeq
Thus the K-theory of the orientifold models described in the previous section
is naturally contained within this two-index set of KR-groups.

To understand what the periodicity \eqn{KRbottper} means physically in terms of
brane charges, we appeal to the ABS construction for Real K-theory. For this,
we define a two-parameter set of Clifford algebras $C_{n,m}$ of the real space
$\real^{n,m}$ as the usual algebra associated with $\real^{n+m}$ together with
an involution generated by the ${\cal I}_m$ involution acting on $\real^{n,m}$.
A Real module over $C_{n,m}$ is then a finite-dimensional representation
together with a $\complex$-antilinear involution which preserves the Clifford
multiplication. The corresponding representation ring $R[spin(n,m)]$ is
naturally isomorphic to the Grothendieck group generated by the irreducible
$\real$-modules $\Delta_{n,m}$ of the Clifford algebra of the space
$\real^n\oplus\real^m$ with quadratic form of Lorentzian signature $(n,m)$. The
ABS map is now the graded ring isomorphism \cite{spingeom,atiyahreal}
\beq
KR(\real^{n,m})~\cong~R\Bigl[spin(n,m)\Bigr]\,/\,R\Bigl[spin(n+1,m)\Bigr]~=~KO^{n-m}(S^0)
\label{absKR}\eeq
This isomorphism relates the groups on the left-hand side of \eqn{absKR} to the
Clifford algebras $C_{n,m}$, so that the $(1,1)$ periodicity \eqn{KRbottper} is
reflected in the $(1,1)$ periodicity of the Clifford algebras,
$C_{n+1,m+1}=C_{n,m}\otimes\real(2)$ \cite{spingeom}.

Let us now consider a $p$-brane of codimension $l=n+m$ in a type-II orientifold
by $\Omega\cdot{\cal I}_m$. The $p$-brane charge is induced by the tachyon
field which is given by Clifford multiplication on the transverse space
$\real^{n,m}$, i.e. $T(x)=\sum_i\Gamma_i\,x^i$ where $\Gamma_i$ are the
generators of the spinor module $\Delta_{n,m}$, and which generates
$\widetilde{KR}(\real^{n,m})$. Under the ABS isomorphism above, this KR-theory
class is multiplied, via the cup product, by the Hopf generator of
$\widetilde{KR}(\complex P^1)=\zed$, or equivalently by the spin bundles which
carry the spinor representation $\Delta_{1,1}$. This gives a class with tachyon
field that generates the KR-group of the new transverse space
$\real^{n+1,m+1}$. This class represents a $p-2$-brane of the type-II
orientifold by $\Omega\cdot{\cal I}_{m+1}$. From this mathematical fact we may
deduce a new descent relation for type-II orientifold theories. A $p-2$-brane
localized at an O$(8-m)^-$-plane in a type-II $\Omega\cdot{\cal I}_{m+1}$
orientifold may be constructed as the tachyonic soliton of a bound state of a
$p$-$\overline{p}$ pair located on top of an O$(9-m)^-$-plane in a type-II
$\Omega\cdot{\cal I}_m$ orientifold. This realizes the branes of a type-II
orientifold as equivariant magnetic monopoles in the worldvolumes of
brane-antibrane pairs of an orientifold with fixed point planes of one higher
dimension. The former orientifold has $2^{m+1}$ O$(8-m)^-$-planes each carrying
RR-charge $-2^{3-m}$, while the latter one has $2^m$ O$(9-m)^-$-planes
of charge $-2^{4-m}$. In a sense, in this process of tachyon condensation the
number of fixed point planes is doubled while their charges are lowered by a
factor of 2 by a combined operation of charge transfer and dimensional
reduction through the orientifold planes. An example is the non-BPS state
consisting of a D5-brane on top of an orientifold 5-plane in the type-IIB
theory \cite{senbps}, which may be constructed via a tachyon condensate from a
pair of D7-$\overline{\rm D7}$ branes on an orientifold 6-plane in the type-IIA
theory. The 8 O$6^-$-planes each carrying charge $-2$ are transfered to the 16
O$5^-$-planes of charge $-1$.

This new sort of internal symmetry among the type-II orientifolds may be
thought of as a type of $T$-duality symmetry acting on the RR-charges of the
orientifold planes, in analogy with the
transformations described in subsection 3.1. By repeated iteration as before,
we obtain an entire hierarchy of novel bound state constructions of D-branes in
various higher dimensions. To describe the field content, however, we must be
careful about identifying the appropriate homotopy of the relevant vacuum
manifolds. The classifying spaces for Real vector bundles are described in
\cite{krclass}. Consider an orientifold of the type-IIB theory, and a set of
brane-antibrane pairs with worldvolume gauge symmetry $U(N)\times U(N)$. The
$U(N)$ gauge group is endowed with its complex conjugation involution whose
fixed point set is the real subgroup $O(N)$. The tachyon field $T$ is
equivariant with respect to the orientifold group, so that
\beq
T(x,-y)=T(x,y)^*
\label{tachyonKR}\eeq
where $(x,y)\in\real^n\oplus\real^m$. It breaks the worldvolume gauge symmetry
down to $U(N)_{\rm diag}$. The relevant homotopy group generated by
\eqn{tachyonKR} comes from decomposing the one-point compactification of
$\real^{n,m}$ into upper and lower hemispheres as described in subsection 2.2,
such that the tachyon field is the transition function on the overlap. The
D-brane charges thereby reside in the KR-group of the transverse space which is
given by
\beq
\widetilde{KR}(\real^{n,m})=\pi_{n,m}\Bigl(U(N)\Bigr)_R
\label{KRhomotopy}\eeq
where the equivariant homotopy group is defined by the maps $S^{n,m}\to U(N)$
which obey the Real equivariance condition \eqn{tachyonKR}. The refined weak
Bott periodicity theorem for stable equivariant homotopy in KR-theory then
reads
\beq
\pi_{n,m}\Bigl(U(N)\Bigr)_R=\pi_{n+1,m+1}\Bigl(U(2N)\Bigr)_R
\label{KRbotthom}\eeq
In a similar way one may relate the Real K-groups
$\widetilde{KR}^{-1}(\real^{n,m})=\widetilde{KR}(\real^{n+1,m})$ to the stable
equivariant homotopy of the complex Grassmanian manifold $U(2N)/[U(N)\times
U(N)]$. In this way we arrive at the Real version of the homotopy sequence
\eqn{botthomotopy}. The first isomorphism from a type-IIB orientifold to a
type-IIA orientifold preserves the structure of the orientifold planes, while
the second step of the sequence decreases the fixed point plane dimension by 1.

This gives a novel generalization of the $T$-duality and descent relations
between IIA and IIB orientifolds, which would be very
interesting to reproduce using string theoretic arguments, as in \cite{senbps}.
Note that the relevance of the Dirac monopole in the orientifolds is not
surprising, given its prominent role in the type-II theories of section 2.
Having identified these orientifold symmetries, we may now restrict our
attention to D-brane charges living in the groups \eqn{KRdiff}, and hence
proceed to the real Bott periodicity relations, as in subsection 3.1. The
sequence of homotopy groups in \eqn{botthomKO} now arises from the fact
\cite{krclass} that if the KR-involution acts freely, then the classifying
space for stable equivariant homotopy reduces to the coset space $U(2N)/O(2N)$.
The rest of the isomorphisms now follow as before. Note that the gauge fields
living on the brane worldvolumes in these cases must also satisfy an
equivariance condition like \eqn{tachyonKR}. A similar analysis can be done for
the $(1,1)$ peridicity of KH-theory \cite{kh} which gives internal relations
among branes localized on O$(9-m)^+$-planes.

\newsubsection{Internal Symmetries}

The Real K-groups are in a certain sense ``universal'' as they contain all of
the other generalized cohomology theories. This feature follows from the many
further internal symmetries present in KR-theory. In particular, there are the
natural periodicity isomorphisms \cite{atiyahreal}
\beq
KR(X\times S^{0,m})=KR^{-2m}(X\times S^{0,m})
\label{KRSper}\eeq
for $m=1,2,4$. The periodicity of KR-theory with coefficients in $S^{0,m}$
follows by using the multiplication in the fields $\real,\complex,\quater$,
respectively, and $(1,1)$ periodicity. Note that for $m=1$ we have the
isomorphism \eqn{KRKiso}, so that \eqn{KRSper} then reduces to the complex Bott
periodicity theorem \eqn{bottKn}, while the $m=4$ case leads immediately to the
real periodicity theorem \eqn{bottRealKn}. In fact, there is the usual
decomposition \cite{atiyahreal}
\beq
KR^{-n}(X\times S^{0,m})=KR^{-n}(X)\oplus KR^{-n+m+1}(X)~~~~~~\forall m\geq3
\label{KRm3split}\eeq

The case $m=2$ is special and the corresponding KR-groups are isomorphic to the
Grothendieck groups generated by the category of self-conjugate vector bundles
over $X$ \cite{atiyahreal}. This self-conjugate K-theory has Bott periodicity
4, which from \eqn{bottrealKn} we see is associated with symplectic Chan-Paton
gauge bundles. This fact has been exploited in \cite{bgh} to relate this type
of K-theory to certain orientifold compactifications of type-II theories
without vector structure \cite{wittenvec}, i.e. those with the same number of
both O$(9-m)^-$ and O$(9-m)^+$ planes of cancelling charge that require no
D-branes in the vacuum configuration, and hence have no gauge group. The
$T$-duality transformations in these theories is also explained in
\cite{bgh}. These transformations, as well as the brane descent relations in
these models, are now natural consequences of the periodicity and homotopy
properties of the KR-groups explained in the previous subsection. It would be
interesting to construct these transformations more explicitly, thus verifying
the results of \cite{bgh}. It would also be interesting to see if the various
duality properties of orientifold compactifications without vector structure
\cite{wittenvec}, which differ somewhat from those described in section 3, can
be deduced from the internal symmetries described in this section.

Thus the internal symmetries of KR-theory encompass most of the brane descent
relations which are based on periodicity theorems, and they further emphasize
the role played by Hopf fiber bundles in the topological classification of
D-branes. In addition, the further symmetry relations in orbifold models may be
thought of as originating within the equivariant structure of KR-theory, and
from the usual Bott periodicities of the given equivariant K-groups, with
solitonic configurations provided by equivariant versions of the four canonical
solitons coming from the Hopf fibrations. It would also be interesting to study
more closely the internal symmetries of type-II orientifolds arising from
quotients by the operator $\klein\cdot{\cal I}_m$, which induce the Hopkins
K-groups $K_\pm(X)$. For instance, as discussed in \cite{gukov}, the products
with the Thom spaces of $\complex$ and $\complex/\zed_2$ induce, respectively,
the periodicity isomorphisms which identify the Hopkins K-groups of
$\real^{n,m}$ with $\real^{n+2,m}$, and of $\real^{n,m}$ with $\real^{n,m+2}$.
The first periodicity builds a $p-2$-brane from a $p$-$\overline{p}$ brane pair
through the usual monopole and leaves the orientifold plane structure
unchanged, while the second one realizes a $p-2$-brane inside a
$p$-$\overline{p}$ pair through a $\zed_2$-equivariant monopole and lowers
the dimension of the orientifold planes by 2. This phenomenon is similar to
that described above for the $\Omega\cdot{\cal I}_m$ orientifolds, and the
monopole symmetries can be seen to naturally arise from the definition of the
Hopkins K-functor as the usual equivariant K-functor (see the right-hand side
of \eqn{Kequivmap}). Again it would be most interesting to carry out these
constructions from a more physical standpoint.

\bigskip

\noindent
{\bf Acknowledgements:} We thank P. Di Vecchia, F. Lizzi, J. M\o ller, R. Nest,
N. Obers, S. Schwede, G. Semenoff and G. Sparano for helpful discussions, and
P. Ho\v rava and G. Landi for comments on the manuscript.

\end{document}